\documentclass[aps,preprintnumbers,superscriptaddress,nofootinbib,aps,prd,floatfix,twocolumn]{revtex4-2}
\pdfoutput=1 
\usepackage{graphicx}%
\usepackage{amsthm} 
\usepackage{amsmath,amssymb,physics}
\usepackage{url} 
\usepackage{lineno} %
\usepackage{xspace}  %
\usepackage{braket}  %
\usepackage{hyperref}
\usepackage{xcolor} 
\usepackage{comment} %
\usepackage{array} %
\usepackage{%
            soul}
\usepackage{longtable}
\usepackage{xparse}
\usepackage{subfigure}

\let\oldcite\cite
\renewcommand{\cite}[1]{\mbox{\oldcite{#1}}}

\NewExpandableDocumentCommand\mcc{O{1}m}
    {\multicolumn{#1}{l}{#2}}
\usepackage{multirow}
\newcolumntype{C}[1]{>{\centering\let\newline\\\arraybackslash\hspace{0pt}}m{#1}}

\definecolor{nicered}{rgb}{0.5,0.,0.}
\definecolor{nicegreen}{rgb}{0.,0.5,0.}
\definecolor{niceblue}{rgb}{0.,0.,0.5}
\definecolor{darkpink}{rgb}{0.8,0.47,0.47}
\hypersetup{colorlinks,citecolor=nicegreen,linkcolor=nicered,urlcolor=niceblue}
\usepackage{enumitem} 
\usepackage[normalem]{ulem}
\setlist{nolistsep} 
 \usepackage{setspace}

\newcommand{\fanto}{{\sf Fant\^omas}\xspace}  %
\newcommand{\meta}{metamorph\xspace}
\newcommand{\xfitter}{{\mbox{\sf xFitter}}\xspace}  %
\newcommand{\FantoPDF}{{\mbox{\sf FantoPDF}}\xspace}  %
\definecolor{darkblue}{rgb}{0.0,0,0.5}
\definecolor{darkgreen}{rgb}{0.0,0.3,0.0}
\definecolor{redish}{rgb}{0.675,0,0.2}
\definecolor{red}{rgb}{0.8,0,0}
\definecolor{green}{rgb}{0,0.6,0}
\definecolor{blue}{rgb}{0,0,0.8}

\newcommand{\orcid}[1]{\,\href{https://orcid.org/#1}{\includegraphics[width=9pt]{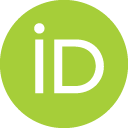}}}

\newcommand{\orcidFO}{0000-0001-6799-2436} %
\newcommand{\orcidPN}{0000-0003-3732-0860} %
\newcommand{\orcidAC}{0000-0001-8906-2440} %
\newcommand{\orcidMAX}{0009-0003-0139-4072} %
\newcommand{\orcidLK}{0009-0007-5639-0350} %

\def\smu{{Department of Physics, Southern Methodist University,
    Dallas, TX 75275-0175, USA \looseness=-1}}
    
\def\unam{{Instituto de F\'isica,
  Universidad Nacional Aut\'onoma de M\'exico, Apartado Postal 20-364,
  01000 Ciudad de M\'exico, Mexico\looseness=-1}}

 \modulolinenumbers[5]

\begin{document}
\preprint{FERMILAB-PUB-23-695-V,SMU-HEP-23-04}
\author{Lucas Kotz\orcid{\orcidLK}}\affiliation{\smu}
\author{Aurore Courtoy\orcid{\orcidAC}}
\email{aurore@fisica.unam.mx}\affiliation{\unam}
\author{Pavel Nadolsky\orcid{\orcidPN}}\email{nadolsky@smu.edu}\affiliation{\smu}
\author{Fredrick Olness\orcid{\orcidFO}}\affiliation{\smu}
\author{Maximiliano Ponce-Chavez\orcid{\orcidMAX}}\affiliation{\unam}

 \date{\today}

\begin{abstract}
We explore the role of parametrizations for nonperturbative QCD functions in global analyses, with a specific application to extending a phenomenological analysis of the parton distribution functions (PDFs) in the charged pion realized in the xFitter fitting framework. The parametrization dependence of PDFs in our pion fits substantially enlarges the uncertainties from the experimental sources estimated in the previous analyses. We systematically explore the parametrization dependence by employing a novel technique to automate generation of polynomial parametrizations for PDFs that makes use of Bézier curves. This technique is implemented in a C++ module Fantômas that is included in the xFitter program.  Our analysis reveals that the sea and gluon distributions in the pion are not well disentangled, even when considering measurements in leading-neutron deep inelastic scattering. For example, the pion PDF solutions with a vanishing gluon and large quark sea are still experimentally allowed, which elevates the importance of ongoing lattice and nonperturbative QCD calculations, together with the planned pion scattering experiments, for conclusive studies of the pion structure.
\end{abstract}

\title{An analysis of parton distributions in a pion with Bézier parametrizations}
\maketitle

\tableofcontents

\newpage
\section{Introduction \label{sec:intro}}
The internal structure of hadrons, as a window on versatile aspects of
strong interactions, is at the center of forefront studies of quantum
chromodynamics (QCD) in the nonperturbative regime.  
New experiments will map internal composition of various hadron
species in three dimensions  
across a broad range of QCD scales, including the
perturbative/non-perturbative transition region~\cite{AbdulKhalek:2021gbh,Quintans:2022utc,Accardi:2023chb}.  
The experimental developments are complemented by vigorous theoretical
efforts to model the hadron structure using nonperturbative
quantum-theoretical methods as well as lattice gauge theory~\cite{Aguilar:2019teb,Ji:2020ect,Abir:2023fpo}. 

Global phenomenological analyses of hadron scattering data serve as a
bridge between experiment and theory. In these approaches,
parametrizations of nonperturbative correlator functions quantifying
the hadron structure are determined from a combination of
hard-scattering experimental data and theoretical inputs. The most
advanced global analyses are dedicated to determination of unpolarized
parton distribution functions  (PDFs) of nucleons and nuclei
\cite{Kovarik:2019xvh}. These analyses include multiloop
hard-scattering contributions, up to the next-to-next-to-leading order
(NNLO) in the QCD coupling strength $\alpha_s$, to assure accuracy and
precision of the resulting PDFs. They also include sophisticated
statistical machinery to evaluate various types of uncertainties.  

Recently, the ever-evolving global analyses have been applied to
study PDFs of pions~\cite{Barry:2018ort, Novikov:2020snp, Cao:2021aci,
 Barry:2021osv} within refined theoretical and methodological
frameworks that supersede the earliest such analyses in
Refs.~\cite{Sutton:1991ay, Gluck:1991ey, Gluck:1999xe}.
In addition, fits of the pion PDFs in specific models have been proposed, 
such as the statistical model~\cite{Bourrely:2022mjf} (and references therein) or the
light-front wavefunction formalism~\cite{Pasquini:2023aaf}. 

Phenomenological studies of pion PDFs draw attention in light
of concurrent rapid developments in lattice QCD that
quantitatively predict characteristics of the PDFs (flavor
composition, Mellin moments, momentum fraction dependence, ...) by
computing matrix elements of bilocal current operators for the PDFs or
related quantities~\cite{Zhang:2018nsy, Izubuchi:2019lyk,
   Joo:2019bzr, Gao:2020ito, Sufian:2019bol, Sufian:2020vzb,Fan:2021bcr,
   Karpie:2021pap,Gao:2022iex,
   Zhang:2023bxs}. 
More broadly, the pion structure, while experimentally less accessible
than the nucleon one, is studied extensively
in QCD theory with the goal to elucidate how the pion's two-body bound state
emerges as a pseudo-Goldstone boson from the dynamical
breaking of chiral symmetry. The pion structure is expected to hold
clues to the symmetry-breaking mechanism responsible for the emergence
of hadron masses,  e.g. \cite{Petrov:1998kg,RuizArriola:2002bp, Polyakov:2009je, Radyushkin:2009zg, Lu:2022cjx}. Such a mechanism may be most obvious for the pions, the lightest long-lived mesons, whose mass decomposition is influenced by the contribution coming from the current quark masses far less than the kaons and heavier mesons. The upcoming experiments will investigate the transition
from this picture, valid at nonperturbative scales of 1~GeV or less,
to another picture in terms of weakly interacting quarks and gluons, applicable
in the multi-GeV energy range, for the pions and mesons in general.

Learning from the experimental observables about the pion's primordial
structure via PDF fits invariably relies on 
quantification of contributing uncertainties, and those arise from
several generic classes of factors: experimental, theoretical,
parametric, and methodological \cite{Kovarik:2019xvh}.
In this article, we address the uncertainty due to the choice of the
functional forms for PDFs at the initial scale of QCD evolution -- the
type of uncertainty that has not been assessed systematically
in the recent pion fits. Potentially significant, the parametrization
uncertainty on the PDFs is not automatically reduced
by increasing the event statistics or the order of the perturbative
calculation. Its assessment requires dedicated techniques,
whether the PDFs are approximated
by polynomials, neural networks, or other functional forms.
In the nucleon global fits, sampling over plausible functional forms
was shown to contribute a large part of the total published uncertainty
\cite{Hou:2019efy,Courtoy:2022ocu}; its effect is especially evident
in the kinematic regions with poor experimental constraints, as
has been demonstrated e.g.\ in the fits utilizing the neural
networks~\cite{Ball:2010de}.  
The estimation of the parametrization uncertainty
is greatly simplified for typical QCD observables that depend only
on a few leading PDF combinations. For such observables, one strives
to learn the likely spread of PDFs from fitted data by repeating the
analysis for a manageable collection of PDF parametrizations.

We have developed a C++ package \fanto to parametrize a variety of
nonperturbative correlator functions using polynomial shapes realized
by B\'ezier curves.  
The B\'ezier parametrizations can approximate an arbitrary continuous
function with desired accuracy as a consequence of the
Stone-Weierstrass approximation theorem. That theorem carries the same significance for the polynomial functions as the universal approximation theorems \cite{cybenko_approximation_1989,hornik_universal_1990,hornik_approximation_1991} do for neural networks. The control parameters  of the B\'ezier curves are
nothing but the values of the PDFs themselves at user-specified
control points. These parameters have a direct interpretation, in
contrast to the free parameters of neural networks and traditional
parametrization approaches. 
Those benefits of a Bernstein polynomial basis were taken advantage of in an early examination of structure functions~\cite{Yndurain:1977wz}.
Today, we know that the B\'ezier curve framework  also serves to mitigate
instabilities in the polynomial interpolation associated 
with the Runge's phenomenon. It aims to make exploration of the
parametrization dependence both interpretable and amenable for
automation.  
It will be fully described in an upcoming publication. 

We incorporated the \fanto package into the public code \xfitter~\cite{xFitterwebsite}  and
employed it to gain insights about the pion PDFs, starting from and
then expanding the recent analysis of pion PDFs published in
\cite{Novikov:2020snp}. We determine these PDFs at the next-to-leading
order (NLO) accuracy by fitting pion-nucleon Drell-Yan (DY) pair
production (largely constraining the quark PDFs), prompt photon production, and newly added data
from HERA leading-neutron production in deeply inelastic scattering
(DIS) (constraining the gluon PDF at $x<0.1$). We fit the PDFs using
several trial functional forms and produce a final ensemble of Hessian
error PDFs by combining the PDFs from the trial fits using the
meta-PDF combination method \cite{Gao:2013bia}. The resulting ensemble
of pion error PDFs is named ``\FantoPDF." 

Exploration of pions, with their simple valence partonic structure and
the central role played by chiral symmetry breaking in their
formation, raises important physics questions that can be elucidated in new
experiments~\cite{Aguilar:2019teb,Arrington:2021biu,AndrieuxSPIN}. How does the pion's valence quark PDF depend on the
partonic momentum fraction $x$? How large is the pion's gluon PDF at
QCD scales of about 1 GeV? Answering these questions requires one to
 thoroughly explore and interpret the PDF parameterization
dependence. The B\'ezier parametrizations are designed for this
purpose.  

For example, the shape of the valence PDFs in the asymptotic limit
$x\to 1$ reflects several factors entering at disparate energy
scales. Dynamical chiral symmetry breaking generically manifests
itself as broadening of the pion distribution amplitude and the pion
quark PDF at a low hadronic scale. Kinematic constraints in the
quasielastic limit lead to a power-law fall-off of structure
functions and PDFs at $x\to 1$ as a function of the number of
spectators and helicity combinations in low-order Feynman
diagrams. Perturbative QCD radiation further suppresses the PDFs at
the largest momentum fractions and high scales as a consequence of
the radiative energy loss from the fastest partons inside the hadron.   

Discrimination among these factors based on the empirical data has
been challenging, not the least because of the parametrization
dependence.  
Various lattice calculations confirm a broad, nearly flat
pion distribution amplitude at low scales~\cite{Holligan:2023rex,
  Gao:2022vyh},  
and a broad valence  pion PDF~\cite{Izubuchi:2019lyk,Joo:2019bzr, 
  Sufian:2019bol,Gao:2020ito,Lin:2020ssv,Alexandrou:2020gxs,Alexandrou:2021mmi}.
These features predicted from  first principles can be compared
against the central PDFs and uncertainty bands of the \FantoPDF
ensemble in lieu of directly comparing against data. The impact of the
parametrization uncertainty on the derived quantities, such as the moments
and logarithmic derivatives of the PDFs, can be examined.  
By trying a variety of flexible functional forms, we suppress the bias
present in the case of using one fixed form. Many functional behaviors
(e.g., given by different polynomials) are consistent with the
discrete data -- this {\it functional mimicry} emphasized in our recent
study \cite{Courtoy:2020fex} stands in the way of determining the
unique analytical form of the PDFs from the discrete data sets. The
functional mimicry also holds a positive aspect, namely that
polynomial functions generated on the fly can realize a wide range of
functional behaviors.

On the side of the data analysis, the available hard-scattering experimental data on the pion have been collected 
in the 1980's and has only a limited constraining power on the PDFs, as has been already emphasized in 1990's
\cite{Sutton:1991ay, Gluck:1991ey, Gluck:1999xe}. Restrictive parametrization forms fail to capture
these limitations when comparing against nonperturbative QCD methods, not to mention the so-far ignored 
uncertainties due to the flavor composition of the quark sea and nuclear PDFs. We make a step toward obtaining 
trusted uncertainty estimates.

Section~\ref{sec:Bezier} introduces the key mathematical formalism of
PDF functional forms based on B\'ezier curves. Section~\ref{sec:PionFitSetup} sets up the characteristics of the global analysis that leads to the \FantoPDF for the pion in Section~\ref{sec:baseline_fanto}. The results are discussed in Section~\ref{sec:results_fanto}. Then we compare the \FantoPDF results with previous global analyses (Section~\ref{sec:GA_comp}) and lattice evaluations (Section~\ref{sec:lattice}), to finally conclude in Section~\ref{sec:Conclusions}. 

\section{Parton distributions parameterized by B\'ezier curves \label{sec:Bezier}}
\subsection{Motivation}
The simplest parameterizations of the hadron PDFs at the initial scale $Q_0$ of the DGLAP evolution 
can be of a ``baseline'' functional form,
\begin{eqnarray}
    x\,f_i(x, Q_0^2)&=& A_i x^{B_i} (1-x)^{C_i} \equiv F_i^{car}(x) \ ,
    \label{eq:param_generic} \label{eq:carrier}
\end{eqnarray}
where $i$ runs over all partons. 
The normalization coefficients $A_i$ are chosen to satisfy the number and momentum sum rules. 
The powers $x^{B_i}$ and $(1-x)^{C_i}$ capture the Taylor series expansions in the limits $x\to 0$ and $1$.
We call this form a {\it carrier function} $F_i^{car}(x)$. 

To introduce more freedom at $x$ away from 0 or 1, we multiply the carrier function $F_i^{car}(x)$ by another function 
$F_i^{mod}(x)$ that we call a {\it modulator}:
\begin{eqnarray}
    x\, f_i(x, Q_0^2)&=& F_i^{car} \times F_i^{mod} \ .
    \label{eq:fanto_FF_0}
    \label{eq:fanto_FF}
\end{eqnarray}
The modulator function is taken to be
\begin{eqnarray}
   F_{i}^{mod}\left(x; {\cal B}^{(N_m)}  \right)
   &=&  
 1+ {\cal B}^{(N_m)}(y) 
     \label{eq:modulator}
\end{eqnarray}
in terms of a polynomial ${\cal B}^{(N_m)}(y)$ of order $N_m$ 
and argument $y\equiv y(x)$, where $y(x)$ is a {\it stretching function} to be defined below. [In the simplest realization, $y=x$.] 
 The baseline parametrization of Eq.~(\ref{eq:param_generic}) is recovered for ${\cal B}^{(N_m)}=0$.

In our methodology, the  ${\cal B}$ polynomial is chosen to be a B\'ezier curve, 
which  has been extensively used in various applications, 
as it provides a flexible functional interpolation. 
A B\'ezier curve is defined as
\begin{equation}
{\cal B}^{(N_m)}(y) = \sum_{l=0}^{N_m} c_l\ B_{N_m,l} (y)
\label{eq:BezierCurve}
\end{equation}
by introducing a  basis of Bernstein polynomials,
\begin{equation}
    B_{N_m,l} (y)\equiv \left(\begin{matrix} N_m  \\ l  \end{matrix}\right) y^l (1-y)^{N_m-l} \ ,
    \label{eq:BNml}
\end{equation}
where the  $\binom{N_m}{l}$ is a binomial coefficient. 

Polynomial functional forms of the type of Eq.~(\ref{eq:fanto_FF_0}) are widely used in global analyses. 
The CTEQ-TEA group 
builds their modulator function from Bernstein polynomials \cite{Hou:2019efy}, 
while the MSHT group employs Chebyshev polynomials \cite{Bailey:2020ooq}.  
Once the degree $N_m$ of the polynomial is fixed, either polynomial basis results in the same modulator. 
The central advantage of the B\'ezier curve is that the numerical coefficients of the polynomial -- $c_l$ in Eq.~(\ref{eq:BezierCurve}) -- 
are computed from the values of the modulator function at a few user-specified values of $x$, or {\it control points} (CPs), 
instead of being fitted directly. These values of the modulator function become the free parameters. 
When the modulator is constructed this way using the B\'ezier curve with $y(x)$ as an argument, we call the whole $xf_i(x,Q_0^2)$ in Eq.~(\ref{eq:fanto_FF_0})  {\it a metamorph} to distinguish it from the other possible forms.

While the B\'ezier representation is algebraically 
equivalent to the traditional polynomial forms, it helps to control numerical correlations among the parameters, which is critical for uncertainty quantification. 
For example, the sign-indefinite orthogonal polynomials 
generally require significant cancellations between coefficients $c_l$ to obtain a positive PDF. This is not an issue with the B\'ezier curves.
Another advantage is that the power $N_m$ of the polynomial can be raised or lowered by adding the control points to the existing parametrization or deleting them. When doing so, the already existing control points are not affected; at this instance, flexibility of the parametrization can be raised or lowered while keeping the input function as is.  

Discrete data do not uniquely fix the degree of a continuous polynomial, reflecting the {\it functional  mimicry} 
of polynomial interpolations explored in Ref.~\cite{Courtoy:2020fex}. 
Functional mimicry is the observation that  polynomials of different degrees 
can yield virtually identical $\chi^2$ over the fitted $x$ range
(leading to $f^{m}(x)=f'^{m'}(x)$ at the fitted points numerically, with $m\neq m'$), while at the same time these polynomials 
produce drastically disparate extrapolations outside of the fitted range. 
Mimicry prevents us from extracting a unique polynomial power 
of the PDFs in an asymptotic $x$ limit (either $x\to 0$ or $1$); 
the phenomenological consequences for the pion PDFs will be discussed in 
Section~\ref{sec:results_fanto}. 

To decouple the flexibility at intermediate $x$ from the asymptotic behaviors at $x\to 0$ or $1$, we introduce a stretching function $y(x)$ such that $F_{i}^{mod}(x)\to 1$ in either asymptotic limit. The other role of $y(x)$ is to aid in the approximation of typical PDF shapes, notably the transition from the approximately logarithmic dependence on $x$ at $x\ll 1$ to the approximately linear one at $x\sim 1$. 

In summary, the \meta construction decomposes $xf_i(x,Q_0^2)$ into a carrier $F_i^{car}$ and  modulator $F_i^{mod}$.
The carrier can be of a standard form in Eq.~(\ref{eq:param_generic}) or another suitable form. 
With a judicious choice of $y(x)$, which we normally do not vary during the fit, the carrier parameters $B_i$ and $C_i$ control the asymptotic limits $x\to 0$ and $1$, respectively. The control points in the modulator determine deviations of the PDF from the carrier at intermediate $x$. The placement of control points can be automated. Correlations between the parameters of the carrier and modulator are reduced. 
In the next subsection, we describe the technical realization of the \meta, including the calculation of $c_l$ coefficients and the stretching function we employ.

\subsection{Technical realization \label{sec:MetamorphRealization}}

\subsubsection{Computing B\'ezier coefficients \label{sec:ComputeCoeff}}
In constructing our polynomials, we will rely on the unisolvence theorem stating that the coefficients of a polynomial ${\cal B}^{(N_m)}(y)$ of degree $N_m$  can be computed exactly given a vector ${\underline P}\equiv {\cal B}^{(N_m)}(y_i)$ for $y_i=y(x_i)$ at $N_m+1$ control points $x_i$ \cite{Courtoy:2020fex}. We utilize a notation in which $S$ (no underline), $\underline V$, and $\underline{\underline M}$ denote a scalar, an $(N_m+1)$-dimensional vector, and an $(N_m+1)\times (N_m+1)$ matrix. Hence, the B\'ezier curve ${\cal B}^{(N_m)}(y)\equiv {\cal B}$ in Eq.~(\ref{eq:BezierCurve}) can be derived in a matrix form in various ways \cite{Farin:2001,pomax}, such as
\begin{equation}
  {\cal  B} =  {\underline Y}^T \cdot \underline{\underline M} \cdot {\underline C}, 
  \label{eq:Bmatrix}
\end{equation}
where $\underline Y \equiv\{y^k\}$ is a vector whose entries are increasing  powers of
$y$ (with $k=0,\ ...,\ N_m$), $\underline{\underline M} $ is an $(\! N_m \!
+ \! 1 \!) \times (\! N_m \! + \! 1 \!)$ matrix  
 containing combinations of binomial coefficients, and $\underline C$
 contains Bernstein polynomial coefficients, $\{c_l\}$.  
 From this relation, we further find that 
\begin{eqnarray}
    {\underline P}&=&{\underline{\underline T}}\, \cdot {\underline{\underline M}}\, \cdot {\underline C},
\end{eqnarray}
 where we introduced a matrix $\underline{\underline T}$ constructed from powers $y^k$ with $k$ ranging between 0 and $N_m$.
 The matrices $\underline{\underline M}$ and $\underline{\underline T}$ are given explicitly in \cite{Courtoy:2020fex}.
The coefficient vector $\underline C$ is then obtained as
\begin{equation}
  \underline C = \underline{\underline M}^{-1} \cdot \underline{\underline T}^{-1} \cdot {\underline P}.
  \label{eq:Cmatrix}
\end{equation}

\subsubsection{Fixed and free control points \label{sec:FixedAndFreeCPs}}

To initialize the \meta function $F_i^{car}(x; A_i, B_i,C_i)\, \left(1+{\cal B}^{(N_m)}(y)\right)$ at the beginning of the fit, we provide the initial coefficients $A_{i,0}$, $B_{i,0}$, $C_{i,0}$ in the carrier and vector $\underline {P_0}$ of the initial polynomial's values at control points ${y_i}$. 
During the fit, we update the carrier parameters as 
\begin{equation}
    B_i = B_{i,0} + \delta B_i, \quad C_i = C_{i,0} + \delta C_i,  
\end{equation}
where the increments $\delta B_i$, $\delta C_i$ are varied by the minimization routine.
We also vary some normalizations $A_i$, while other normalizations are computed using the sum rules.
As for the components $P_i$ of ${\underline P}$ in the modulator, we vary them in a similar way ($P_i = P_{i,0} + \delta P_i$) at {\it free} control points or keep them constant ($P_i = P_{i,0}$) at {\it fixed} control points. For instance, for a $j$-th control point we may decide to keep $P_j=0$, so that the whole $xf(x,Q^2)$ reduces to the carrier at $x=x_j$. 

We can increment the polynomial degree $N_m$ by inserting a control point at some $x=x_{N_m+1}$. As the initial $P_{N_m+1,0}$ at the new point, we can either choose ${\cal B}^{(N_m)}(y_{N_m+1})$ to retain the modulator value from the $N_m$-th degree polynomial, or set $P_{N_m+1,0}$ to any other value, for example, $P_{N_m+1,0}=0$.

\subsubsection{The stretching function \label{sec:sx}}
In the current implementation, we choose a stretching function of the form $y=(g(x))^{\alpha_x}$, with $\alpha_x>0$ and with
\begin{equation}
 g(x)\equiv \frac{1}{\left( \frac{1}{x^6+x_{\rm min}^6} + \frac{1-x_{\rm max}^6}{x_{\rm max}^6}\right)^{1/6}}
\end{equation}
dependent on two constant parameters, $x_{\rm min}$ placed to the far left from the fitted interval of $x$, and $x_{\rm max}$ placed to the far right. It is easy to see that $g(x)\to x_{\rm min}$ when $x \ll x_{\rm min}$, and similarly $g(x)\to x_{\rm max}$ when $x \gg x_{\rm max}$. For $x_{\rm min} \lesssim x \lesssim x_{\rm max}$, $g(x)$ essentially coincides with $x$. The purpose of $g(x)$ is to turn off smoothly the modulator in the asymptotic intervals $0 < x < x_{\rm min}$ and $x_{\rm max} < x < 1$. Indeed, we can place two fixed control points with $P_0 = P_1 = 0$ at $x_0=x_{\rm min}$ and $x_1=x_{\rm max}$. This choice forces $xf_i(x,Q^2)=F_i^{car}(x)$ at $x \ll x_{\rm min}$ and $ x \gg x_{\rm max}$, i.e. the PDF is asymptotically pinned to the carrier. 

In the intermediate $x$ range, we can further rescale $y$ with respect to $x$ using the $\alpha_x$ power to redistribute the flexibility of the polynomial over the $x$ subranges. A possible choice $\alpha_x=0.3-0.5$ helps to capture the fast growth of a PDF at $x<0.1$ and slower dependence at $x>0.1$. 

\begin{figure*}[bth]
     \centering
     \includegraphics[height=3in]{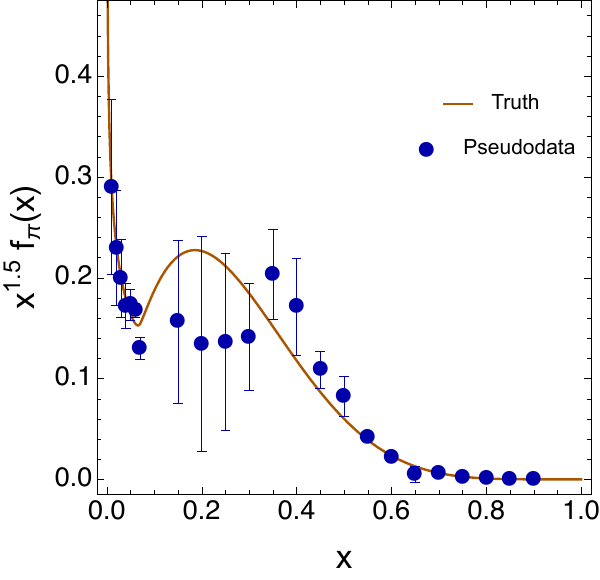}\quad
     \includegraphics[height=3in]{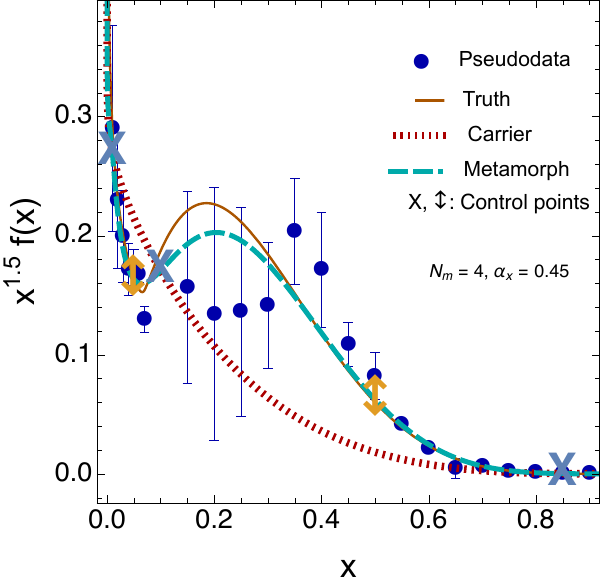}
\caption{Illustration of a fit by a metamorph. (a) An ensemble of data points is generated by random fluctuations around a ``truth'' function constructed from piecewise polynomial interpolations. (b) After minimization of the $\chi^2$ between data and theory, the
  carrier function (short-dashed red curve) has varied and the
  values of all control points have been shifted, helped by the
  modulator, {\it i.e.}, the B\'ezier curve. The ``fixed'' CPs (blue
  crosses) lie on the updated carrier function. The free CPs (golden arrows) can deviate from it. The final result is
  the long-dashed cyan curve, labeled ``Metamorph". It better approximates the data than the carrier only. This example is
  given for $N_m=4, \,\alpha_x=0.45$.}
\label{fig:toyfit1}
\end{figure*}

\subsection{An illustration of aleatory and epistemic uncertainties \label{sec:ToyIllustration}}
To demonstrate versatile features of the \meta approach and to validate its {\sf C++} implementation, we implemented it in a {\sf
  Mathematica} notebook~\cite{Maxthesis:2022} and tested it first by reproducing the coefficients of the Bernstein polynomials in CT18 NNLO parametrizations \cite{Hou:2019efy} by fitting the whole $xf_i(x,Q_0^2)$ by metamorphs. 

Figure~\ref{fig:toyfit1} illustrates these features in a fit of generated fluctuating one-dimensional data in {\sf Mathematica}.  
In Fig.~\ref{fig:toyfit1}(a),
we generated such data by random fluctuations around a ``truth'' function (solid ocher curve), which in this specific example is given by a union of several piecewise polynomial curves. The goal is to fit the data with the \meta
methodology and compare it with the (known) truth. If the fitted function consists only of the carrier as in Eq.~(\ref{eq:param_generic}), the fit generally has a difficulty in fitting dip-bump structures like the one seen at $x=0.05-0.3$. Therefore, after fitting the data with the carrier only as a first estimate, we attach a modulator using two fixed control points at 
$x_{\rm min}=10^{-3}$ and $x_{\rm max}=0.9$, which are indicated by crosses in Fig.~\ref{fig:toyfit1}(b). %

We then have freedom to choose, in accordance with the size and shape of the data~\cite{Kovarik:2019xvh},  the degree $N_m$ of the modulator polynomial, the $x$ positions of $N_m+1$ control points, and the stretching parameter $\alpha_x$. We also can vary the style of the fit by keeping some parameters free or fixed. For example, the carrier parameters $B_i$ and $C_i$ could be held constant just like the preprocessing factors adopted by NNPDF \cite{Ball:2016spl}, or they could be fitted together with the modulator, as other groups do.

In the example in the right subfigure, we selected $N_m=4$, $\alpha_x=0.45$ and added two variable control points at $x=0.05$ and 0.5 (indicated by the arrows), as well as a fixed control point that pins $xf(x)$ to the carrier at $x=0.1$. We varied both the carrier and control points and obtained a best-fit \meta curve (long-dashed cyan), which is the product of the updated carrier function (short-dashed red curve) and the B\'ezier modulator curve. 

\begin{figure}[bth]
\includegraphics[width=0.9\columnwidth]{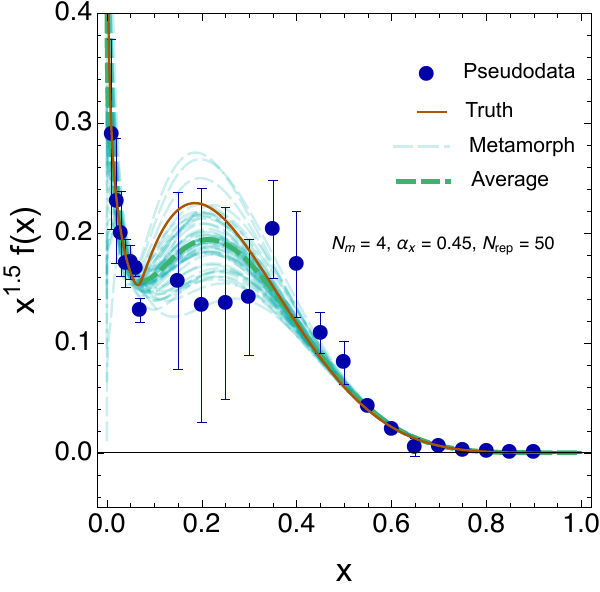}  
\includegraphics[width=0.9\columnwidth]{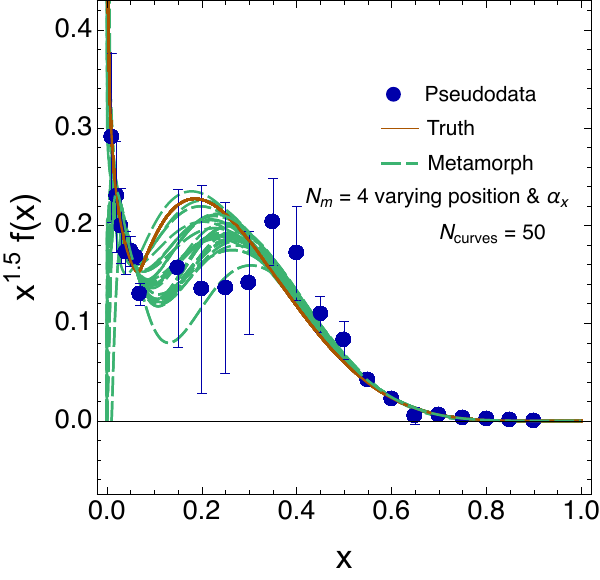}
\caption{The \fanto technique illustrated by applying the bootstrap (or importance) sampling on the data (upper plot) or the F\^antomas methodology, that consists in sampling over representative choices for the CPs and the scaling factor $\alpha_x$ (lower plot). The resulting uncertainties are displayed in cyan (bootstrap) and green (parameter-space sampling).}
\label{fig:meta_boot_fanto}
\end{figure}

The ultimate purpose for the \meta methodology is to streamline quantification
of uncertainties. Once a central fit has been determined,
say, the long-dashed cyan curve of Fig.~\ref{fig:toyfit1}, its
full statistical meaning is obtained through the propagation of the
two classes of uncertainties: the aleatory and epistemic
ones. The aleatory class in this case propagates from stochastic fluctuations of data. In
the upper plot of Fig.~\ref{fig:meta_boot_fanto}, we estimate the aleatory uncertainty using the bootstrap
method, one of the possible error propagation techniques. Also known
as resampling or importance sampling, it involves generating $N_{\rm rep}=50$
replicas of the same data set by fluctuating the central data values 
according to their respective standard deviations. Each
replica of the data is fitted by a respective \meta (a light cyan curve in
the upper plot of Fig.~\ref{fig:meta_boot_fanto}); their (unweighted)
average is plotted here in green. The curves obtained after
bootstrapping all correspond to the same \meta settings (here $N_m=4,\
\alpha_x=0.45$, unvaried positions of CPs).  
The curves are superimposed on one replica of the data considered earlier.
While each curve gives a good fit to their corresponding fluctuated replica, they at the same time give 
increasingly worse fits to the rest of the replicas when the number of parameters increases. 
This loss of generalizability of the PDF models can be partly mitigated by cross-validation (not done in
our toy example). A single bootstrapped replica does not present a good fit to the averaged data, 
while their mean (the green line) usually does \cite{Hou:2016sho}.

To estimate the epistemic uncertainties for some replica of data, it is necessary to sample
over the space of models, which in the case of the metamorphs means the 
sampling over the settings to investigate a large collection of polynomials~\cite{Courtoy:2022ocu}. 
The lower plot in Fig.~\ref{fig:meta_boot_fanto} shows a bundle of $N_{\rm curves}=50$ \meta fits done on the same replica of data, now varying the positions and types of control points, as well as the stretching power $\alpha_x$. The degree of the polynomial, here set to $N_m=4$, 
can also be varied. The bundles in the upper and lower plots are not the same: bootstrapping of the data does not sample over the settings of the model and analysis. The total uncertainty can be estimated by combining the curves from both sources, which can be done using the METAPDF \cite{Gao:2013bia} or another method for combination of the PDF uncertainties. We will use this strategy to combine the experimental (aleatory) and parametrization (epistemic) uncertainties in the analysis of real pion data in the subsequent sections. 

\subsection{Implementation of \meta in \xfitter}
In parallel with the {\sf Mathematica} implementation, we have developed a C++ package \fanto 
with the \meta parametrizations, which was then incorporated into the  
\xfitter fitting program~\cite{Alekhin:2014irh,xFitterwebsite,Buckley:2014ana,Botje:2010ay}. 
Just like the other standard parameterizations available in the PDF library {\sf pdfparams} of \xfitter, the \meta functions can be used for any flavor of choice. 

In contrast to the standard parametrizations, the \meta  parameterization requires the user 
to provide positions and types of control points, as well as parameters of the carrier and the stretching functions. All these parameters are stored in a dedicated steering card file, which is sufficient for reproducing the \meta parameterizations between the fitting runs. The steering card also records the initial values of the modulator at the free control points. To simplify manipulations with the control points, \fanto treats the shifts from the initial values at control points as free parameters. The shifts are varied to minimize the $\chi^2$. An updated \fanto steering card is produced at the end of each fitting run.  

Several options in the \fanto module control the
flexibility of the metamorphs. One option 
allows the carrier function, Eq.~(\ref{eq:carrier}), to be fixed
($\delta B_i=\delta C_i=0$) or varied during the minimization
process.  
An initial guess for the carrier parameters can be obtained by running an $N_m=0$ fit  first, from
which the obtained best-fit carrier parameters can be retained while increasing the
degree of ${\cal B}^{(N_m)}$. 

An alternative strategy, which we employed in some candidate fits of the pion PDFs, 
is to allow free carrier parameters and fix $F^{mod}(x)=1$ at $0 < x \ll x_{\rm min}$ and $x_{\rm max} < x < 1$, as explained in Sec.~\ref{sec:sx}. Guidelines on the usage of the \fanto environment in \xfitter will be given in a
separate publication that will accompany the public release of the {\sf C++} module and {\sf Mathematica} notebook~\cite{FantoTech}.

\subsection{Placing of control points and Runge's phenomenon \label{sec:Runge}}
Control points are a crucial aspect of the \meta not just because their positions $x_i$
can leverage the space of solutions by spanning more functional forms. 
Strategic placing of the control points prevents the instability in the 
interpolation when using a high-degree polynomial, also known as the Runge's phenomenon.
In Eq.~(\ref{eq:Cmatrix}), the instability arises when matrix
$\underline{\underline T}$, the only part that depends on spacing of control points in $y$, 
has a vanishing determinant. This tends to happen, {\it e.g.}, with equidistant spacing of control points 
and $N_m$ above 5-10, or when two control points are too close. 

The \fanto code computes the condition number for $\underline{\underline T}$ according to 
the Frobenius matrix norm. Together with a small ${\rm det}\,\underline{\underline T}$, a large 
condition number is another indicator that flags the likely polynomial instability. Users should
seek to minimize this metric by setting up a \textit{well-behaved} matrix
$\underline{\underline T}$. This is achieved by taking advantage of the \meta parameters, 
\textit{e.g.} the stretching power $\alpha_x$.  

With the typical choices we tested, the B\'ezier curve representation stabilizes interpolation 
with $N_m$ as high as 15-20. Such range of $N_m$ is entirely sufficient for the current phenomenology, as even  the most precise nucleon PDF fits like CT18 and MSHT20 introduce at most about 30 effective degrees of freedom divided among all PDF flavors. 

\section{Setup of the global analysis of the pion PDFs}
\label{sec:PionFitSetup}

\subsection{Pion PDFs within the B\'ezier framework}
\label{sec:BezierPion}

In contrast to the extensively studied proton PDFs,  
only a  few PDF fits have investigated 
the charged pion PDFs. The most recent analyses of this kind are done with modern fitting
frameworks, e.g.\ developed by JAM (in a Monte-Carlo framework with
Bayesian statistics \cite{Barry:2018ort,Barry:2021osv, Cao:2021aci}) and \xfitter (in  a Hessian framework \cite{Novikov:2020snp}) groups. 
These investigations employed fixed polynomial forms with few parameters and reported negligible improvements 
in $\chi^2$, the figure of merit, when trying more flexible functional forms.

In anticipation that there may be significant parametrization dependence,  
the following sections extend the \xfitter framework 
to include more general parametrizations for the initial PDFs using B\'ezier curves. 
Flexible functional forms are necessary to ensure  proper sampling of the full space of solutions; 
this has been  shown to affect uncertainty quantification for large-dimensional spaces~\cite{Courtoy:2022ocu}. 
While the parameter space for the pion
PDF is arguably smaller than that of the proton,   the extrapolation
regions are spanned differently; gaps in the data coverage at small ($x\ll 0.1$)
and intermediate ($x\sim 0.1$) momentum fractions require PDF parametrizations permitting enough variation 
within the gaps, not only at $x\to 0$ or $1$. Bootstrapping with fixed functional forms alone does not capture the extra uncertainty in the gap regions, as was observed e.g.\ in studies with hybrid fitting frameworks even for a small data ensemble~\cite{Bacchetta:2012ty,Benel:2019mcq}. 

We work at the next-to-leading order (NLO) in the QCD coupling strength $\alpha_s$ and assume the same flavor composition of pion PDFs at the initial scale $Q_0^2 =1.9\mbox{ GeV}^2$ as in the \xfitter study \cite{Novikov:2020snp}. In addition to the gluon PDF $g(x)$, we introduce the total valence and sea quark distributions, $V(x)$ and $S(x)$, and determine the PDFs for individual flavors on the assumptions that the PDFs for the two constituent (anti)quarks are the same in $\pi^+$ and $\pi^-$, and the light quark sea is flavor-blind:
\begin{eqnarray}
&u^{\pi^+}=\bar d^{\pi^+}=\bar{u}^{\pi^-}=d^{\pi^-} = V/2  + S/6, \label{v}\\
& \bar{u}^{\pi^+}=\bar{d}^{\pi^-}=s^{\pi^{\pm}}={\bar s}^{\pi^{\pm}} = S/6.
\label{S}
\end{eqnarray}
Note that {\it a priori} $V(x)$ is not positive-definite. 
Using the above symmetries, we can compute all PDFs from three independent distributions, ${q=\{V, S, g\}}$, with each given by a \meta functional form
\begin{equation}
    x\,f_i(x, Q_0^2)= A_i x^{B_i} (1-x)^{C_i} \left[ 1+ {\cal B}^{(N_m)}(y(x)) \right].
    \label{xfpion} 
\end{equation}

Among the three PDF normalizations, $A_V$ is fixed by the valence sum rule.
Depending on the candidate fit, we choose either $A_S$ or $A_g$ to be independent
and find the third normalization from the momentum sum rule for $\langle x f \rangle =\int_0^1 xf(x)\ dx$,
\begin{equation}
    \langle x V \rangle + \langle x S \rangle + \langle x g \rangle = 1.
    \label{momsum}
\end{equation}
In particular, we find that the gluon momentum fraction $\langle xg\rangle$ can be zero according to the data. Exploring such solutions with the \xfitter program is most feasible when we choose $A_g$, rather than $A_S$, to be independent.

The \meta parametrizations in Eq.~(\ref{xfpion}) can be compared against the ones in the \xfitter study \cite{Novikov:2020snp}:
\begin{eqnarray}    x\, V(x, Q_0^2)&=& A_V\, x^{B_V} (1-x)^{C_V},\nonumber\\
    x\, S(x, Q_0^2)&=& \frac{A_S}{B\left(B_S+1, C_S+1\right)}\, x^{B_S}
    (1-x)^{C_S},\nonumber\\ 
    x\, g(x, Q_0^2)&=& A_g\,(1+C_g)\,  (1-x)^{C_g}.
    \label{eq:param_xfitter}
\end{eqnarray}
Here, the Euler-Legendre beta function $B(B_S+1, C_S+1)$ equates the momentum fraction $\langle xS\rangle$ of the sea quarks directly to $A_S$, taken to be independent.

At $Q$ scales above the charm and bottom masses, $m_c= 1.43$ GeV and $m_b=4.50$ GeV, the charm and bottom PDFs are included in the DIS coefficient functions with mass dependence using the modified Thorne-Roberts scheme \cite{Thorne:2006qt}.

\subsection{Selection of data \label{sec:Data}}
 \begin{figure}[t]
\begin{center}
    \includegraphics[width=\columnwidth]{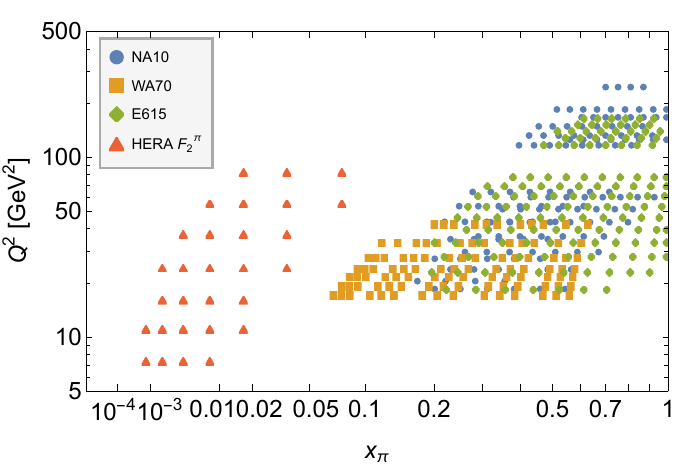}
    \caption{Kinematic coverage in $x_{\pi}$ and $Q^2$ of the Drell-Yan, prompt photon and leading-neutron DIS data used in this work.}
    \label{fig:xQ2}
\end{center}
\end{figure}

The \xfitter pion PDFs were determined
 from the analysis of Drell-Yan (DY) pair production by $\pi^-$ scattering on a tungsten target by
 E615 (140 data points)~\cite{Conway:1989fs} and NA10 (140 data points)~\cite{NA10:1985ibr}, as well as
 prompt-photon ($\gamma$) production in $\pi^-$ and  $\pi^+$ scattering on a tungsten target by WA70 (99 data points)~\cite{WA70:1987bai}. [See Ref.~\cite{Novikov:2020snp} for details on the data selection.] The kinematical coverage of those data is most sensitive to the 
 pion PDFs at large $x$ values, as shown in Fig.~\ref{fig:xQ2}. DY
 processes are most sensitive to valence distributions, with the gluon contributing mainly through the DGLAP evolution. The prompt-photon data
 provide additional constraints on the gluon distribution above
 $x\gtrsim 0.1$. 

The data coverage did not allow the \xfitter analysis to separate the quark sea ($S$) and the gluon ($g$) PDFs at $x<0.1$, as will be shown clearly in Sec.~\ref{sec:Validation}. To remedy the entanglement of the sea and gluon
distributions, we followed the proposal of the JAM collaboration to
analyze pion DIS data from leading-neutron (LN)
processes~\cite{McKenney:2015xis, Barry:2018ort}. %
 In such processes,
the initial-state pion emerges with a relatively low virtuality in the proton-to-neutron 
transition. A flux factor modeling that $\pi PN$ transition is convoluted with the DIS structure function $F_2^\pi$
that is of interest here. %
However, theoretical uncertainties remain large in the extraction of $F_2^\pi$, with the H1 and ZEUS collaborations adopting different extraction procedures.%

 This idea was originally explored at HERA~\cite{ZEUS:2002gig,
   H1:2010hym, H1:2010udv}. The kinematic variables for the DIS process include
 $x_{Bj}=Q^2/ (2p\cdot q)$, 
the Bjorken scaling variable for the proton target, the momentum
fraction carried by the leading neutron, $x_L$, and the
 momentum transfer between the proton and the neutron, $t$. 
Then,
the momentum fraction corresponding to DIS on a pion target
with 4-momentum $(1-x_L)p$ is given by 
\begin{eqnarray}
    x_{\pi}%
    ={x}/{(1-x_L)}.\nonumber
\end{eqnarray}
The  pion flux factor has been evaluated in various models (see, {\it
  e.g.}, Refs.~\cite{Holtmann:1994rs,Hobbs:2014fya, McKenney:2015xis}
for ampler discussions).\\ 

In the present analysis, we follow the 
prescription and selection of data from the H1 collaboration~\cite{H1:2010hym} for a minimal inclusion of LN data.  
The H1 analysis identifies the single-pion exchange approximation
({\it e.g.}, close to the pion pole) to be valid for $x_L>0.7$ and
estimates that LN production could be used to extract the pion PDF
in the range $0.68<x_L<0.77$ and at low momentum transfer.
 This kinematical region
 corresponds to $29$~LN data points illustrated by red
 triangles in Fig.~\ref{fig:xQ2}. The H1 collaboration quotes an integrated pion flux factor, $\Gamma_{\pi}$, 
\begin{eqnarray}
    \Gamma_{\pi}(x_L)&\equiv&  f_{\pi N}(x_L)%
    =\int_{t_0(x_L, p_T)}^{t_{min}(x_L)}\, dt\, f_{\pi N}(x_L, t)\nonumber
\end{eqnarray}
from the
 light-cone representation of Ref.~\cite{Holtmann:1994rs}, which, in the
 selected kinematics, gives 
\begin{eqnarray}
    f_{\pi N}(x_L=0.73)
    &\simeq& 0.13\pm 0.04\,.
    \label{eq:flux_xL073}
\end{eqnarray}
In Ref.~\cite{H1:2010hym},
an uncorrelated uncertainty of 30\% is quoted to encompass variations due to different expressions for the pion flux. We adopt the same prescription. For simplicity, we do not separate the 5\% normalization uncertainty from the 30\% uncertainty, as it is negligible in the total uncertainty compared to the pion flux uncertainty.

For the chosen H1 data, the overall structure function  can be expressed as a product
\begin{eqnarray}
    F_2^{LN(3)}(Q^2, x, x_L=0.73)&=& f_{\pi N}(x_L=0.73)\, F_2^{\pi} (x_{\pi}, Q^2)\,,\nonumber\\
    \label{eq:pionSF}
\end{eqnarray}
from which we can extract the pion DIS structure function, $F_2^{\pi} (x_{\pi}, Q^2)$. Then, $29$ data points are added to the \xfitter data library in the form of a $F_2$ structure function.
\\

\subsection{Validation of the fitting methodology \label{sec:Validation}}

\begin{figure}%
    \centering   
\includegraphics[width=0.9\columnwidth]{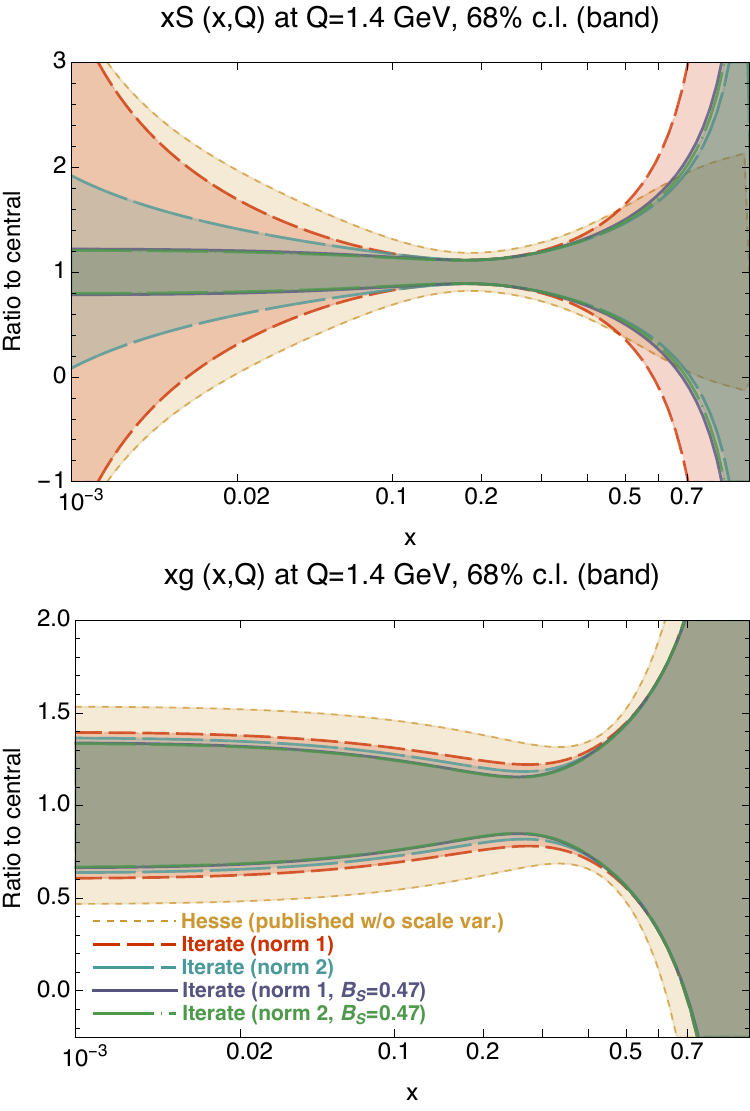} 
    \caption{Ratios of the uncertainties to the respective central PDFs for the sea $xS(x,Q)$ (upper panel) and gluon $g(x,Q)$ (lower panel): \xfitter published
      results~\cite{Novikov:2020snp} using {\sf Hesse} evaluation of the uncertainties
      (without uncertainty from scale
      variations) in mustard; the same ``norm 1'' fit  with {\sf Iterate}
      uncertainties in medium-dashed red; and ``norm 2'' fit with  
      {\sf Iterate} uncertainties in long-dashed cyan. Then, for a fixed $B_S=0.47$,
      ``norm 1'' and ``norm 2'' fits are compared as overlapping dark blue and
      green curves. 
      } 
    \label{fig:xfitterfits_sg}
\end{figure}

Once the \fanto environment has been set up using an independent C++ package, we have tested it against
the published results of \xfitter for the pion
PDF~\cite{Novikov:2020snp}. For this initial comparison, we excluded the LN DIS data -- a new component introduced in our fit.

When using the lowest-order polynomials ($N_m=0$), the parametrizations
in Eqs.~(\ref{eq:fanto_FF})  and~(\ref{eq:param_xfitter}) differ only by their
normalization, which we shall call ``norm~1" and ``norm~2." 
In the case of \xfitter parameterization ``norm 1,'' the normalizations $A_{V, S, g}$
directly determine the respective momentum fractions. The potential downside of the Euler beta function 
in the denominator of $xS(x,Q^2_0)$ is that it introduces a nonlinear relation among $A_S$, $B_S$, and $C_S$ when calculating the uncertainty in the Hessian approach. 

On the other hand, the parameters $A_i$ in the \meta parametrization do not directly
represent the momentum fraction of each flavor -- that's ``norm~2." 
To recover the results obtained with Eqs.~(\ref{eq:param_xfitter}),
the \meta degree is set to $N_m=0$, which requires one fixed control
point for each PDF flavor $i$. With only the DY and prompt-photon data included, 
phenomenological separation of the sea and gluon distributions mainly relies on the choice of
functional form and propagation on uncertainty into the extrapolation region, which in this
specific case extends starting from $x\lesssim 0.1$ until the end point $x \sim
0$ for the whole $Q^2$ coverage.  
Hence,  the small-$x$ behavior of the sea balances that of the gluon
in a subtle way. The parameter $B_S$ that dominates the small-$x$ behavior of the sea
is poorly determined, and even compatible with zero. We find that the propagation of
the uncertainty following  ``norm~1" or ``norm~2" does not render a similar result.

\begin{figure}[tb]
\begin{center}
    \includegraphics[width=0.9\columnwidth]{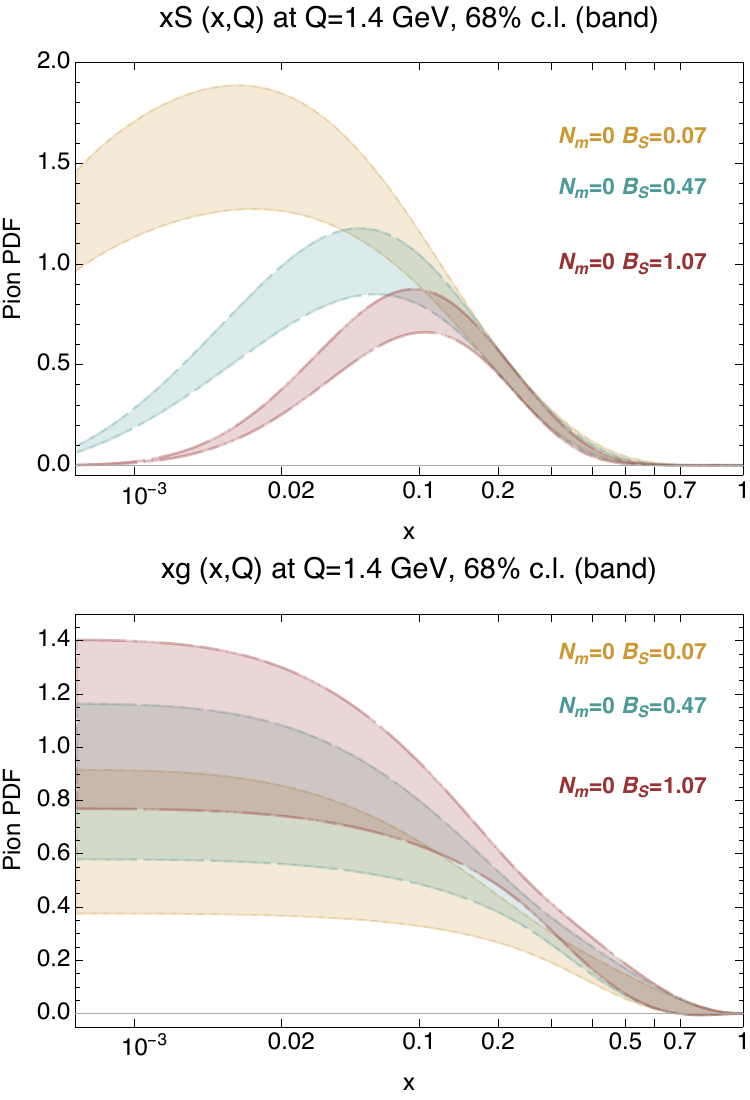}
    \caption{The \fanto PDFs for $N_m=0$, based on DY+$\gamma$ data,
      shown for various values of the fixed parameter $B_S$. 
      The upper panel (lower panel) shows the quark sea (the gluon) PDF and its uncertainty.} 
    \label{fig:Nm0_compar_Bs}
\end{center}
\end{figure}

\begin{table*}[tb]
\def\hs{\hspace{3.5mm}} %
\begin{tabular}%
    {{|l | c | C{2.5cm} C{2.5cm} C{2.5cm}|}}
        \hline
    \hline
 $N_m=0$ (DY+$\gamma$)  & $\chi^2\, [d.o.f=379-5]$  & \(\langle xV \rangle\)& \(\langle xS \rangle\)& \(\langle xg \rangle\) \\ [0.5ex] 
 \hline\hline      
\rule{0pt}{4ex} 
$B_S=0.07$ & $445.70$ & $0.556$ & $0.268$ & $0.177$ \\
\rule{0pt}{4ex}
$B_S=0.27$&$445.38$ & $0.557$ & $0.239$ & $0.204$ \\
\rule{0pt}{4ex}
$B_S=0.47$& $445.29$ & $0.558$ & $0.217$ & $0.225$ \\
\rule{0pt}{4ex}
$B_S=0.67$& $445.36$ & $0.559$ & $0.199$ & $0.243$ \\
\rule{0pt}{4ex}
$B_S=0.87$& $445.52$ & $0.559$ & $0.184$ & $0.257$ \\
\rule{0pt}{4ex}
$B_S=1.27$& $445.76$ & $0.559$ & $0.172$ & $0.269$ \\
  \hline      
\end{tabular}
\caption{Momentum fractions for $N_m=0$ and DY+$\gamma$ data only. Correspond to fits shown in Fig.~\ref{fig:Nm0_compar_Bs} at $Q^2=1.9$ GeV$^2$.
\label{tab:bs}
} 
\end{table*}

To explore the causes of such differences, we have considered two
modalities for the {\sf Minuit} diagonalization of the Hessian
matrix in \xfitter, namely {\sf Hesse} and {\sf Iterate}. 
The former is the default setting, while the latter comes from the
CTEQ collaboration~\cite{Pumplin:2000vx}. 
The 2020 \xfitter study calculated the PDF uncertainties with the {\sf
  Hesse} method. Figure~\ref{fig:xfitterfits_sg} illustrates the
differences between the outcomes of the two methods.\footnote{The technical details will
be shared elsewhere~\cite{FantoTech}. Similar observations have been
recently discussed in Ref.~\cite{Glazov_xFitter_2305}.}
While the best-fit PDFs agree, the {\sf Hesse} error bands are larger than the 
{\sf Iterate} ones. The instability in the Hessian diagonalization can be mended 
by fixing the small-$x$ power $B_S$ of the
sea distribution. Thus, in Fig.~\ref{fig:xfitterfits_sg}
we compare error bands with a free $B_S$ and {\sf Hesse} and {\sf Iterate} methods with ``norm~1'' and ``norm~2,'' 
as well with a constant $B_S=0.47$ and  {\sf Iterate} with ``norm~1'' and ``norm~2.''
With $B_S$ fixed, both the {\sf Hesse} and {\sf Iterate} error bands agree much better
for all three flavors and normalization types.\footnote{Plots for valence distribution are
omitted unless visually relevant.} Practically the same best-fit
parameters and errors are found with all choices, as well 
$\chi^2/N_{\rm dof}=445.3/(379-5)=1.19$ at the best fit.

Poor determination of the small-$x$
behavior of the gluon with free $B_S$ leads to a confusion about the allowed
values for the momentum fraction of the sea. The parameter $A_S$,
quoted as $0.216 \pm 0.025$~\cite{Novikov:2020snp},
translates into $A_S=6\pm 1$ for ``norm 2" with fixed $B_S$ but
delivers a much larger uncertainty, $A_S=6\pm 10$, when
$B_S$ is free to vary.  

After reproducing the  nominal \xfitter results using the new \fanto module with $N_m=0$, a first extension  naturally comes from varying the {\it fixed} value of the small-$x$ exponent of the
sea distribution within some interval, which we take to be $B_S\in [0.07, 1.07]$. 
The \meta parameterizations are still used with 
\(N_m=0\) for each flavor, i.e. the PDFs are given by their respective carrier functions in Eq.~(\ref{eq:carrier}). In each of these fits, the valence PDF converges to the same function, whereas the
sea and gluon PDFs change, with their magnitudes adjusting inversely to one another at $x<0.3$ as \(B_S\) is varied
[Fig.~\ref{fig:Nm0_compar_Bs}]. The corresponding momentum fractions
are listed in Table~\ref{tab:bs}. These results, combined with the
\(\chi^2\) results, indicate that the sea and gluon PDFs are not well
constrained at low $x$. Analogous quantitative results for a DY-only analysis were reached 
by the JAM collaboration~\cite{Barry:2018ort}. 

The individual fits for $B_S= \{0.07,\ 0.27$, $0.47,\ 0.67$, $0.87,\ 1.07\}$ also realize a Lagrange
Multiplier scan over that parameter. The scan indicates a very shallow $\chi^2$ parabola  (see the $\chi^2$
column of Table~\ref{tab:bs}), which, in turn, indicates that the
statistical loss from varying \(B_S\) is negligible. 

From this point on, we will develop the full \fanto
formalism within the \xfitter fitting framework. By adding the LN DIS data at small $x$, we reduce, although not entirely eliminate, the small-$x$ degeneracy of the quark sea and gluon PDFs that was just discussed.
This largely independent 
analysis with a different methodology will be compared again with the
\xfitter 2020 results in Sec.~\ref{sec:GA_comp}. 

\begin{figure}[bht]
\begin{center}
\includegraphics[width = .95\columnwidth]{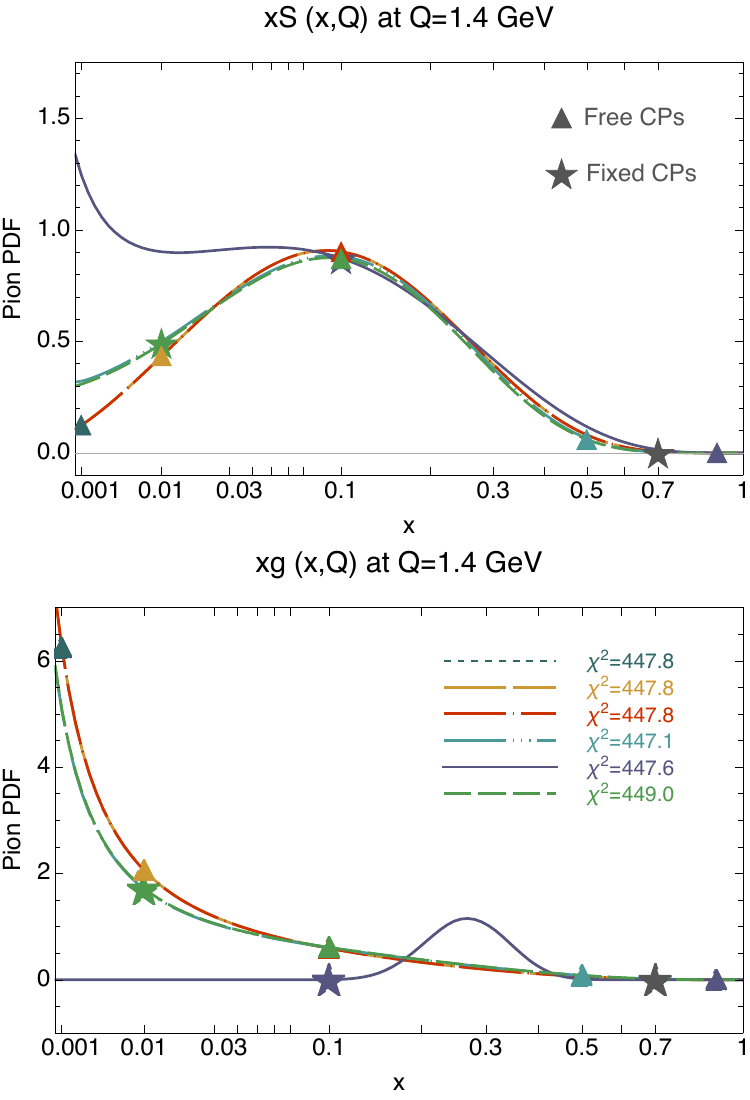}
\caption{The sea and gluon pion PDFs obtained with the \fanto environment for an $N_m=1$
  and $\alpha_x=0.75$ setting, shown at $Q_0$. The control points are
  indicated for each curve with a star and a triangle for a fixed and
  free CP, respectively.  Notice that all but two curves share a fixed
  CP at $x=0.7$.}
\label{fig:Nm1_sg}
\end{center}
\end{figure}

\section{The baseline Fant\^omas fit}
\label{sec:fanto_full}
\label{sec:baseline_fanto}

\subsection{Exploration of the parametrization dependence \label{sec:param}}
Building upon a broader and more diverse data set illustrated in
Fig.~\ref{fig:xQ2}, we proceed with an in-depth investigation
of functional form dependencies enabled by the \fanto methodology. 
In the fits to the full data set, we start with $N_m=0$ parametrizations for all flavors 
$i{=}\{V, S, g\}$ and free carrier parameters. We then vary $N_m$ between 0 and 3 independently for each flavor by adding or removing free or fixed CPs. Recall that the number of CPs is kept equal to $N_m+1$, so that they uniquely determine the modulator polynomial. The shape of the best-fit PDF is controlled by the carrier parameters $A_i,\ B_i,\ C_i$, as well as by the degree $N_m$, stretching power $\alpha_x$, and placement of fixed CPs in the modulator. Positions of free CPs, on the other hand, do not modify the best-fit polynomial, given its uniqueness, and rather can be placed according to convenience (e.g., since the function values at CPs tabulate the PDF at user-chosen positions $x_i$).  Altogether, the presented fits have at least 5 and at most 13 free parameters.

Figure~\ref{fig:Nm1_sg} illustrates the variety of solutions obtained by
selecting positions of the fixed and free CPs for $(N_m=1,
\alpha_x=0.75)$. Using the same carrier and fixed CPs, but moving the
abscissa of the free control point, we obtain very similar resulting
sets with equally good $\chi^2$ values -- see the first three labeled
curves of Fig.~\ref{fig:Nm1_sg}.  
For other choices of the fixed control points, the gluon PDF differs in shape 
and can even get close to zero at $Q_0$,
yet always staying non-negative. The latter solution also shows that the 
gluon has an additional freedom at $x\sim 0.3$, where experimental constraints are weak, 
as indicated by a bump in the fifth curve. The relative paucity of the pion data 
allows such solutions, as the \fanto analysis reveals.  
The shown solutions have nearly the same $\chi^2$, all at least as good as the $\chi^2$ of the 2020 \xfitter fit, cf. Table~\ref{tab:bs}, 
despite having more data in the current study. The differences among the solutions in Fig.~\ref{fig:Nm1_sg} 
enlarge the spread in the respective momentum fractions.

\begin{figure}[bht]
\begin{center}
\includegraphics[width = .95\columnwidth]{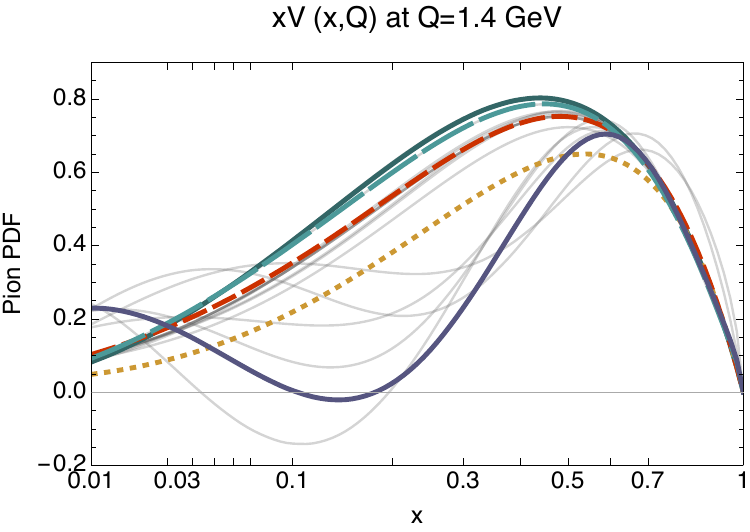}
\includegraphics[width = .95\columnwidth]{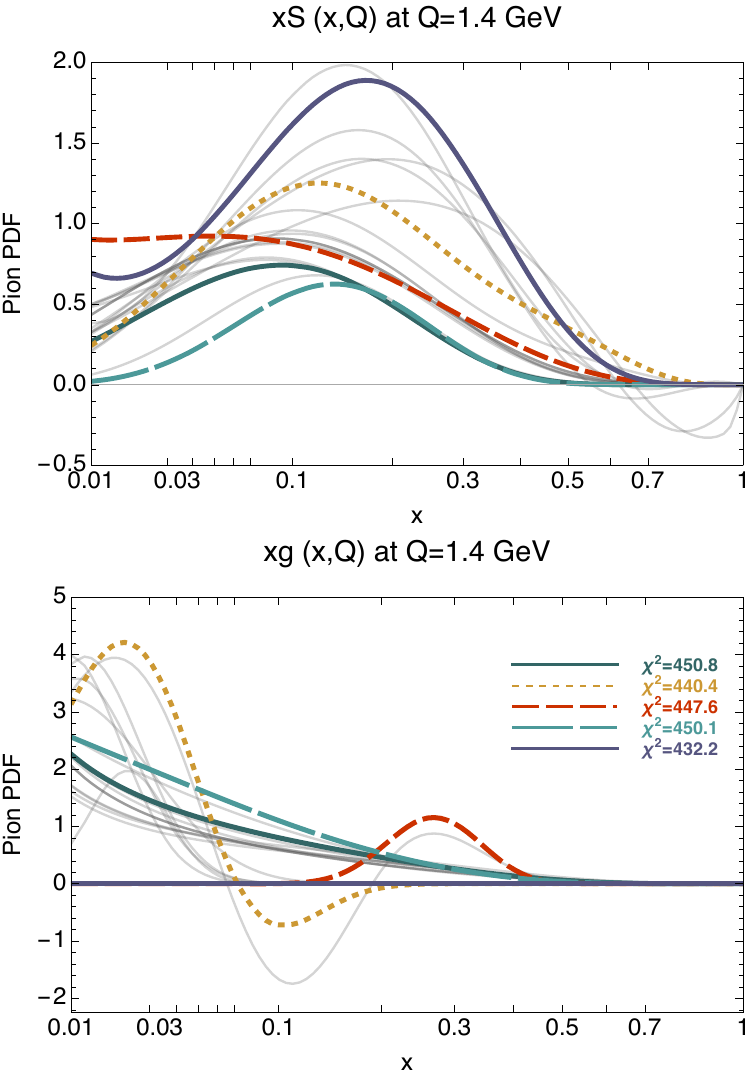}
\caption{Central values for pion PDFs obtained with the \fanto
  environment for various $N_m$, $\alpha_x$ and CP settings, shown at
  $Q_0=\sqrt{1.9}$ GeV. The baseline fits chosen for the final combination are shown in color. Light gray curves illustrate the many possible shapes that were obtained.} 
\label{fig:funky_sg}
\label{fig:combi_CV}
\end{center}
\end{figure}

Expanding to higher degrees of the polynomial, whether across all or some
flavors, revealed an even richer spectrum of potential
solutions. Among these, the lowest $\chi^2$ value observed in our limited sample was 
$414$, corresponding to a negative sea distribution at large-$x$ values. 
Figure~\ref{fig:funky_sg} presents a subset of about $\sim 100$ such configurations.
In color, we display the fits that are kept for the final combination, for which the fit with the lowest chi-square value ($\chi^2=432.2$) corresponds to a zero-gluon distribution. The second-best $\chi^2=440.4$ belongs to a configuration
that exhibits a negative gluon PDF around $x\simeq 0.1$.
The expected variance of the chi-square for $N_{\rm pts}=408$ and
varying $N_{\rm par}$ between 7 and 13 amounts to $\delta
\chi^2=\sqrt{2(N_{\rm pts}-N_{\rm par})}\simeq 30$ at the $1\sigma$
level~\cite{Kovarik:2019xvh}. We retain fits with $\chi^2\lesssim
450$, a few units below the variance of $\chi^2_{\rm min}=432$.  Most of the \meta sets exhibit chi-square values well within
$1\sigma$ from the minimum value that was obtained. 

The spread of the solutions in Fig.~\ref{fig:funky_sg} quantifies the parametrization dependence that contributes to the epistemic uncertainty. As we reminded in Sec.~\ref{sec:ToyIllustration} using a toy example, the epistemic uncertainty (corresponding to the lower panel in Fig.~\ref{fig:meta_boot_fanto}) is independent from the aleatory one. The latter could be either estimated by
resampling/bootstrapping the data $N_{\rm rep}$ times, as illustrated in the upper panel of Fig.~\ref{fig:meta_boot_fanto} and followed by JAM, or alternatively by diagonalization of the Hessian matrix \cite{Pumplin:2001ct}, as done in \xfitter.
The 68\% c.l. aleatory uncertainties quoted by \xfitter correspond to a variation of
$\Delta \chi^2=1$ around the minimum of a single fit, which has long been known to lead to error bands that are excessively narrow. 
To these, we should add the epistemic uncertainty, the core rationale behind our investigation. 

\subsection{Combination of fits with different parametrizations \label{sec:META}}
\begin{figure}[bht]
\begin{center}
\includegraphics[width = .95\columnwidth]{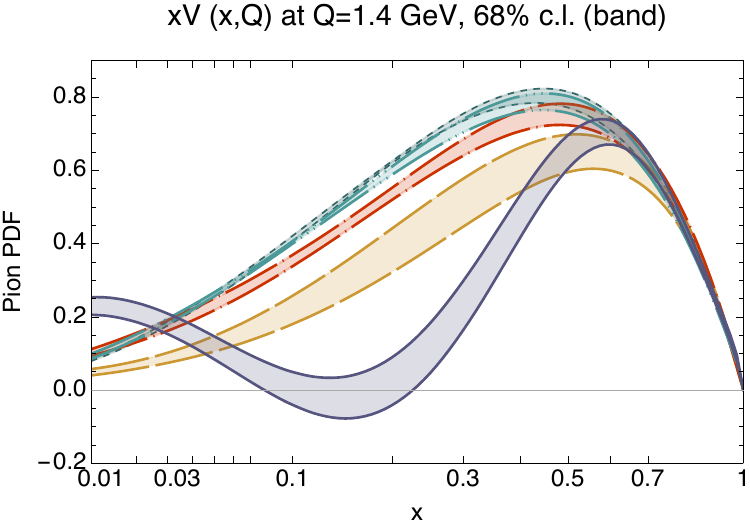}
\includegraphics[width = .95\columnwidth]{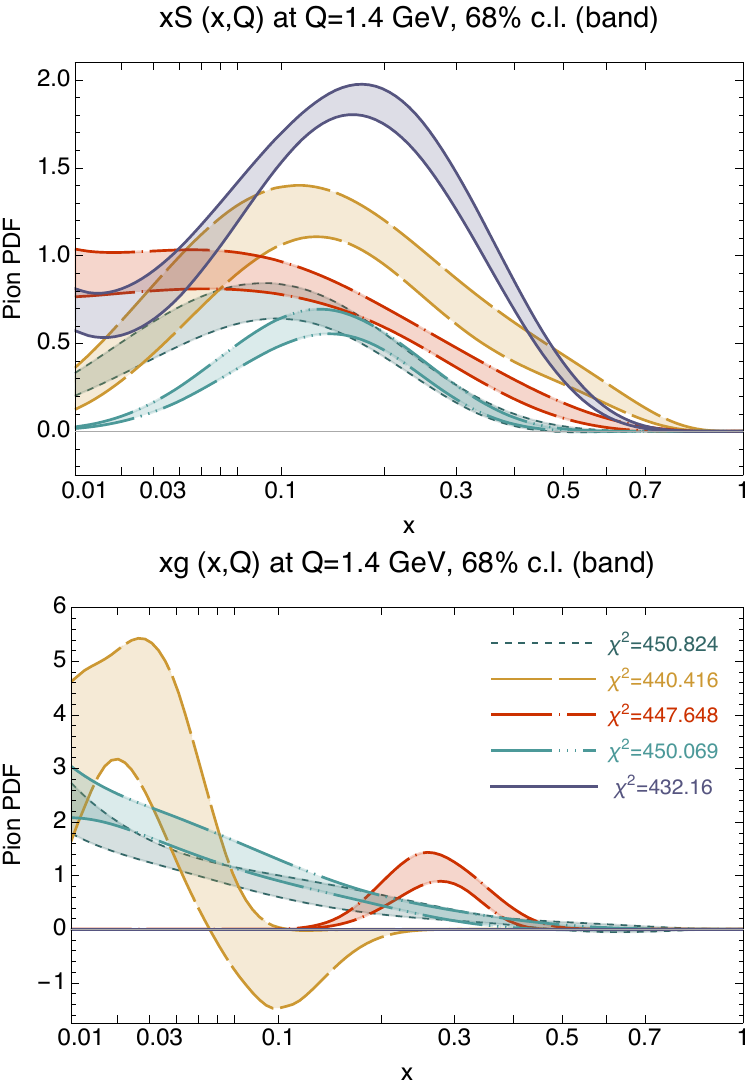}
\caption{Five final sets of the \fanto PDF combination, shown at
  $Q_0=\sqrt{1.9}$ GeV, with their respective Hessian uncertainty band as obtained from the {\sf xFitter} framework.} 
\label{fig:5_Hessian_errorband}
\end{center}
\end{figure}

 In many practical applications with parsimonious parametric dependence, a useful estimate of epistemic uncertainties may be possible through sufficiently representative sampling of acceptable PDF models~\cite{Courtoy:2022ocu}. We follow this paradigm by first selecting five out of 100 explored configurations with good 
 $\chi^2$ so that their central fits cover a broad span in the PDF magnitudes; Fig.~\ref{fig:funky_sg} shows the central PDFs for five such selected parametrizations in color. 
 Appendix~\ref{sec:AppendixData} explores their agreement with individual data sets.
 For each select form, we generate an uncertainty band with $\Delta \chi^2=1$
 using the Hessian method to estimate their aleatory uncertainties. 
 Figure~\ref{fig:5_Hessian_errorband} displays these five error bands. 
 From them, we generate a combined PDF error ensemble using the {\sf METAPDF} method \cite{Gao:2013bia}.  We fully present the final PDFs in the next section.
 
 This procedure is analogous to the PDF4LHC21 combination~\cite{PDF4LHCWorkingGroup:2022cjn}.
It is more direct than introducing either the global or dynamic tolerance, while operationally it serves the same purpose of accounting for the parametrization uncertainty as done, e.g., by introducing the tolerance $\Delta \chi^2>1$ in the CTEQ-TEA analyses \cite{Hou:2019efy}. In extrapolation regions with no data or first-principle constraints, the PDF combination techniques avoid unrealistically small uncertainty that are common with the customary tolerance prescriptions. Tensions among the fitted data sets -- the other rationale for increasing the tolerance in global fits \cite{Kovarik:2019xvh,Pumplin:2001ct,Pumplin:2002vw} -- are not relevant in this pion fit. 

 The combined ensemble is produced using the {\sf mp4lhc} package~\cite{Gao:2013bia}. 
 From each of the five input Hessian ensembles in Fig.~\ref{fig:5_Hessian_errorband}, 
 with $2N_{\rm eig}$ eigenvector PDFs each, we generate 100 Monte-Carlo replicas by following a linear  sampling procedure and symmetric errors~\cite{Hou:2016sho}. For each value of $k\in 1,\cdots, 100$, 
\begin{eqnarray}
&&f^{(k)}(x, Q_0)\label{eq:dXkAsym}\\
&=& f_0 (x, Q_0)\pm  \sum_{i=1}^{N_{\rm eig}}R_{i}^{(k)}f_{i}(x, Q_0) + \Delta,\nonumber
\end{eqnarray}
where $f_{i}$ is the Hessian error set associated with the  displacement along the $i$-th eigenvector, $f_{0}$ the central value, and $\{R_i\}$ is a set of random numbers, each sampled independently from a standard normal distribution in the interval $-\infty < R_i < +\infty$.
A small shift $\Delta$ was applied to all MC replicas in accord with \cite{Hou:2016sho} so that the central value of the resulting Monte Carlo set reproduces then the central value of the input  Hessian set. Given a small number of PDF parameters, $100$ replicas appear sufficient for the reproduction of the Hessian probability, which is assumed to be Gaussian.
Then the five sets of 100 Monte-Carlo replica form the combined  final \fanto Monte-Carlo ensemble, for which we can evaluate the average PDF, {\it e.g.} displayed in Fig.~\ref{fig:fin_together}. 
The original ensembles are not weighted, as
the differences in their numbers of degrees of freedom are negligible with respect to the number of data points
$N_{\rm pts}$. 

When deciding on the selection of the PDF solutions for the final combination, we could not avoid consideration of the prior constraints, such as smoothness of PDFs and positivity of cross sections. In this study, smoothness is achieved by using polynomials with low $N_m$ and sufficient spacing of control points to suppress the Runge phenomenon (cf. Sec.~\ref{sec:Runge}). Notably, the pion data do not disfavor and even prefer the quark and sea gluon PDFs that go negative in certain $x$ intervals at the initial scale $Q_0^2 = 1.9\mbox{ GeV}^2$  or have local extrema. For example, this feature applies to the negative gluon shown by a short-dashed mustard curve in Fig.~\ref{fig:funky_sg}. JAM collaboration has also found such solutions~\cite{Cao:2021aci}. We, however, choose to discard strongly negative PDFs, some of which are displayed in light gray. We include some solutions with weakly negative PDFs to capture the enhanced uncertainty in the gluon and sea sectors at $(x\simeq 0.1, Q\gtrsim Q_0)$ and in other extrapolation regions that must be further constrained by future data. 

The positivity of distribution functions, as a first-principle argument, has been a topic of various research publications~\cite{Candido:2020yat, Collins:2021vke, Candido:2023ujx}. Hadron-level cross sections  must be non-negative 
everywhere in $x$ and $Q^2$. Whether this necessitates non-negative PDFs in perturbative QCD depends on other factors, including the order and scheme of the calculation, mixing of scattering channels, all-order resummations and a variety of nonperturbative contributions [see a general discussion in Ref.~\cite{Courtoy:2020fex} and Ref.~\cite{Aicher:2010cb, Barry:2021osv} for threshold resummation]. Furthermore, QCD evolution turns mildly negative PDFs at $Q_0$ and small $x$ into positive ones at slightly higher $Q$ values where perturbative QCD dominates. Requiring the PDFs to be non-negative at small $Q_0$, where perturbative QCD ceases to converge, may artificially suppress the PDF uncertainty at larger $Q$, where the positivity is no longer an issue. In fact, the issue of positivity could be entirely avoided for such PDFs by choosing a slightly higher $Q_0$ to start with, at the expense of a slightly higher $\chi^2$ -- we have checked that it is the case for the sets that display a negative gluon at $x\to 1$. Dependence of pion PDFs on $Q_0$ must be explored in further studies, as it is known to be relevant since the early model evaluations both for the PDFs and exclusive form factors, e.g.\ in ~\cite{Polyakov:2009je, Radyushkin:2009zg}.

To recap, the five fits that constitute the final \FantoPDF ensemble are highlighted with bold colors and fonts on Fig.~\ref{fig:funky_sg}. They are chosen for their distinct shapes, thereby maximally spanning the parameter space explored across the $\sim 100$ \meta configurations that have been tested. Among these selected outcomes, the dashed-mustard curve  corresponds to a \meta configuration yielding a slightly negative gluon distribution 
in the $x\simeq 0.1$ extrapolation region. These fits are combined using the {\sf METAPDF} technique that captures the bulk of the explored parametrization uncertainty, even though it does not include some extreme baseline solutions with negative PDFs, with some shown in grey in Fig.~\ref{fig:funky_sg}.

\section{The \fanto parton distributions of the pion}
\label{sec:results_fanto}

\begin{figure}[b]
\begin{center}
\includegraphics[width = .975\columnwidth]{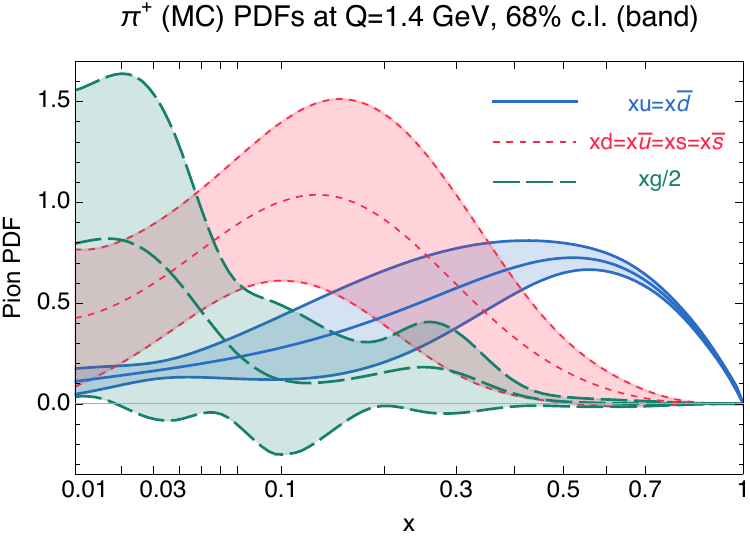}
\caption{The \FantoPDF sets at $Q_0$, shown for a $\pi^+$ flavor configuration.}
\label{fig:fin_together}
\end{center}
\end{figure}

Figure~\ref{fig:fin_together} displays the combined \fanto pion PDF ensemble, constructed from the 
five fits for which the chi-square values of the central PDF
 range between $\chi^2/N_{\rm pts}$ of  1.06 (432/408) and 1.10 (450/408), improving upon 
 the \xfitter's original DY+$\gamma$ fit quality.\footnote{Note that the results have been updated since Ref.~\cite{Courtoy:2023bme}.} 
 A description of the data comparison with the \FantoPDF sets is given in Appendix~\ref{sec:AppendixData}.
 
The statistical meaning of the \fanto ensemble differs from the one expected from resampling (importance sampling), as it
it covers both types of uncertainties (aleatory and epistemic) illustrated in the two panels of Fig.~\ref{fig:meta_boot_fanto}. 
The aleatory uncertainties are estimated by the Hessian error bands %
with $\Delta \chi^2=1$ for the unfluctuated (published) data for each of the five PDF fits that sample the epistemic uncertainty. 
To apprehend the epistemic uncertainty, the sampling of the PDF models should be sufficiently dense in multidimensional space of parameters, as discussed in \cite{Courtoy:2022ocu} and references therein. While sampling becomes a daunting problem in space of many dimensions, its efficiency can be drastically improved in specific applications by optimizing a quantity called {\it a confounding correlation} or {\it data defect correlation} \cite{Hickernell:2018a,Meng:2022dec}. Thus, the five fits reasonably capture the spread of PDFs from about a hundred candidate fits obtained in the first step of our study. The likely distances between the true PDFs and their fitted parametrizations are thus determined by the interplay of the confounding correlations with the data quantity and aleatory uncertainties of the data itself. 

\begin{figure}[b]
\begin{center}
    \includegraphics[width=0.9\columnwidth]{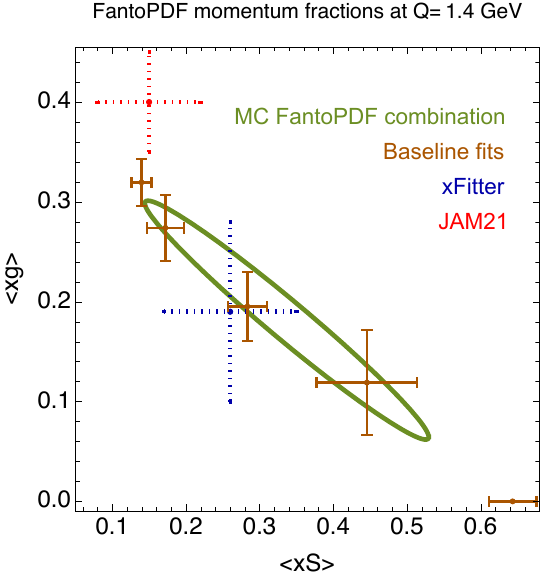}
    \caption{Momentum fractions for the sea and gluon PDFs obtained in the five baseline fits (orange-brown) and the MC \FantoPDF combination (green ellipse) at $Q_0$. JAM21 and \xfitter results, to be discussed in Sec.~\ref{sec:GA_comp}, are displayed in red and blue, respectively. }
    \label{fig:moment_frac_corr}
\end{center}
\end{figure}

From this ensemble of pion PDFs with comparable $\chi^2$, we can compute the momentum fractions $\langle xf\rangle$ integrated over the whole range of $x$. The valence sum rule governs the momentum fraction for $V(x,Q_0)$, while the momentum sum rule relates the sea and gluon momentum fractions. Figure~\ref{fig:moment_frac_corr} illustrates the latter at scale $Q_0$, with variations in $\langle xS\rangle$ showing a strong correlation with $\langle xg\rangle$. Here, we plot the momentum fractions for the five baseline fits as well as for the final \FantoPDF combination. We also superimpose the respective momentum fractions from JAM'21 \cite{Barry:2021osv} and \xfitter \cite{Novikov:2020snp} analyses -- they will be discussed in detail in the next section. 

\begin{figure}[bt]
\begin{center}
    \includegraphics[width=0.995\columnwidth]{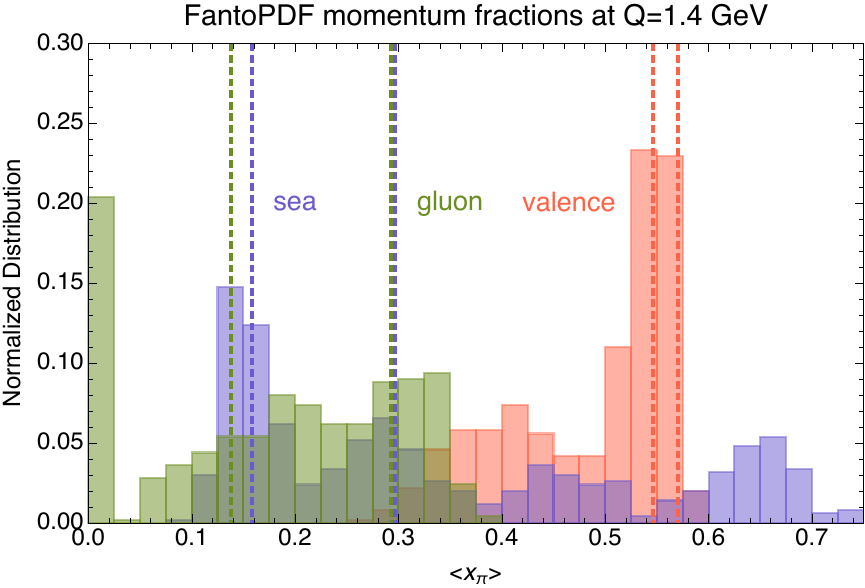}
    \caption{The histograms represent momentum fractions for the valence (red), gluon (green) and sea (blue) PDFs from 500 MC \FantoPDF distributions generated from five candidate fits. The histograms are not symmetric as a consequence of parametrization dependence. Vertical boundaries represent the extrema of momentum fractions for pre-\fanto fits with DY+$\gamma$ data only (Fig.~\ref{fig:Nm0_compar_Bs}).  These results are at the initial scale $Q_0$. }
    \label{fig:moment_frac}
\end{center}
\end{figure}

The picture that emerges is that the flexibility of the functional form strongly affects conclusions about the experimentally allowed momentum fractions by increasing their ranges compared to the fits with fixed parametrization forms. We can further illustrate this point by examining the histograms in Fig.~\ref{fig:moment_frac} for momentum fractions for $\braket{xV}$, $\braket{xS}$, and $\braket{xg}$ obtained with 500 replicas from the five baseline sets. The parametrization dependence in the DY+$\gamma$+LN fit substantially widens the histograms as compared to the full intervals (vertical dashed lines) from the DY+$\gamma$ fit with the fixed functional form in Sec.~\ref{sec:Validation}. In particular, the addition of leading-neutron data only partially improves the separation 
between sea and gluon distributions compared to the pre-\fanto analysis of the DY-only data
[Fig.~\ref{fig:Nm0_compar_Bs}]. Addition of the LN data is not by itself sufficient for narrowing the interval of the allowed sea and gluon momentum fractions. The long $\braket{xS}$-$\braket{xg}$ correlation ellipse for the final \fanto combination in Fig.~\ref{fig:moment_frac_corr} supports this assessment, which is somewhat in a contradiction with the JAM analyses~\cite{Barry:2018ort} -- see Sec.~\ref{sec:GA_comp}. 

While we allow marginally negative fits, the final results produce positive momentum fractions for all components at $Q_0$. 
Table~\ref{tab:mmts_q0} (upper row) summarizes the  results for the momentum fractions at $Q_0$. The $q+\bar{q}$ and gluon momentum fractions
are given at $Q=2$ GeV in Table~\ref{tab:mmts}  (upper row).
\\

\begin{figure}[bt]
\begin{center}
    \includegraphics[width=0.95\columnwidth]{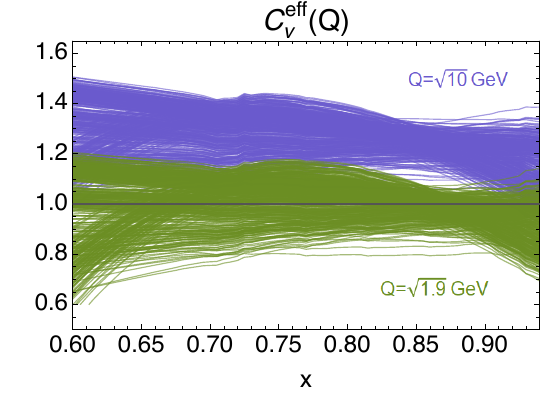}
    \caption{ The effective $(1-x)$ exponent of the valence PDF in the \FantoPDF
      ensemble -- the definition is given in~\cite{Courtoy:2020fex}. In green, the
      effective exponent at $Q_0=\sqrt{1.9}$ GeV and, in blue, at
      $\sqrt{10}$ GeV. The plot is cut at $x=0.94$ for
      grid-extrapolation reasons. We have verified analytically that
      the  highlighted B\'ezier curves of Fig.~\ref{fig:funky_sg}
      converge to $C_V^{\mbox{\tiny eff}}=1$ at most for $x\to 1$ at $Q_0$.} 
    \label{fig:a2eff}
\end{center}
\end{figure}

We conclude this section by reviewing our findings for the pion valence PDF in the limit $x\to 1$.
Recent debates on manifestations of nonperturbative dynamics in high-energy data
made the large-$x$ behavior of the pion quark PDF a go-to 
topic~\cite{Aguilar:2019teb, Arrington:2021biu,AbdulKhalek:2021gbh,Barry:2021osv}. 
The quark-counting rules -- an early prediction reflecting the kinematic constraints in the quasielastic region --
suggest a $(1-x)^{\beta=2}$ fall-off of the pion quark PDFs when $x\to 1$. This expectation does not
account for many dynamic contributions at either low or
large momentum fractions that affect interpretation of realistic measurements~\cite{Courtoy:2020fex}. 
In this regard, the  present phenomenological analysis does not qualitatively differ from the previous
 recent ones: the fall-off of the valence PDF at large $x$ is compatible with $\beta=C_V^{\mbox{\tiny eff}}=1$ at
 $Q_0=\sqrt{1.9}$ GeV (see Fig.~\ref{fig:a2eff}), in spite of the
 multiple functional forms that have been considered (Fig.~\ref{fig:funky_sg}).

\section{Comparison with other  global analyses}
\label{sec:GA_comp}

Now that we have established the power of the \meta parametrization in
the \fanto formalism, we proceed to further comparisons of our results to previous
extractions of the pion PDFs at NLO.  The fitted data sets vary among the
fitting groups. 
While the \fanto analysis augments the full DY and
prompt photon data set of \xfitter with $29$ LN DIS data points, JAM authors
choose to fit a selection of DY data for invariant masses lower than the $\Upsilon$
mass (noticeable around $\mu^2=100$ GeV$^2$ in
Fig.~\ref{fig:xQ2}), leading to a reduced DY set {\it w.r.t.} \fanto's. No prompt-photon data are used in JAM's analysis. Also, their analyses are based on a larger pool of LN DIS data,  stemming in turn
from a combined study of the E866 proton
$\bar{d}-\bar{u}$ and the HERA (H1 and ZEUS) LN DIS in the context of a
pion-cloud approximation~\cite{McKenney:2015xis}. Hence, our analyses differ
 in the distributions of data points per process. JAM has similar numbers of data point for DY and LN. For \fanto, the majority of the data is from DY pair and prompt-photon production. 
\begin{figure}[t]
\begin{center}
\includegraphics[width = .95\columnwidth]{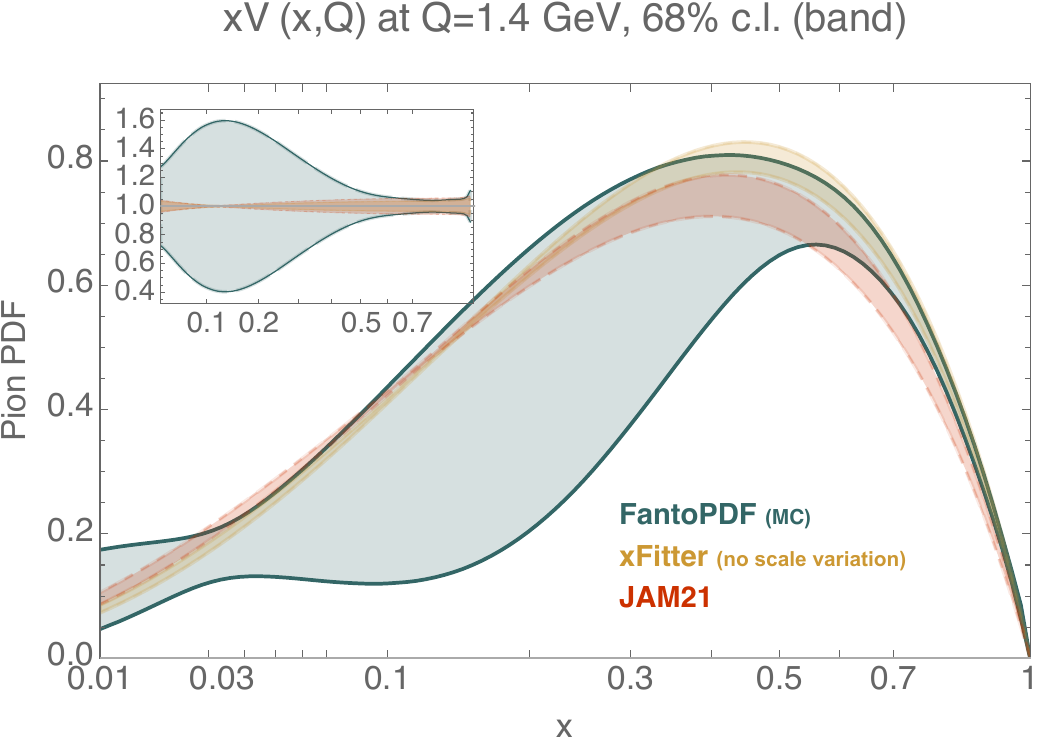}
\caption{The valence PDF of the final \fanto ensemble compared to JAM21 and \xfitter results,
  at $Q=1.4$ GeV. For the \FantoPDF set, the
  $68\%$ CL of the MC output is shown. \xfitter's results are plotted
  without accounting for the uncertainty coming from the scale
  variation. The inner frame shows the ratio to the central value of
  each set -- symmetric uncertainties are used for all three sets.} 
\label{fig:fin_v}
\end{center}
\end{figure}
\begin{figure}[tb]
\begin{center}
\includegraphics[width = .95\columnwidth]{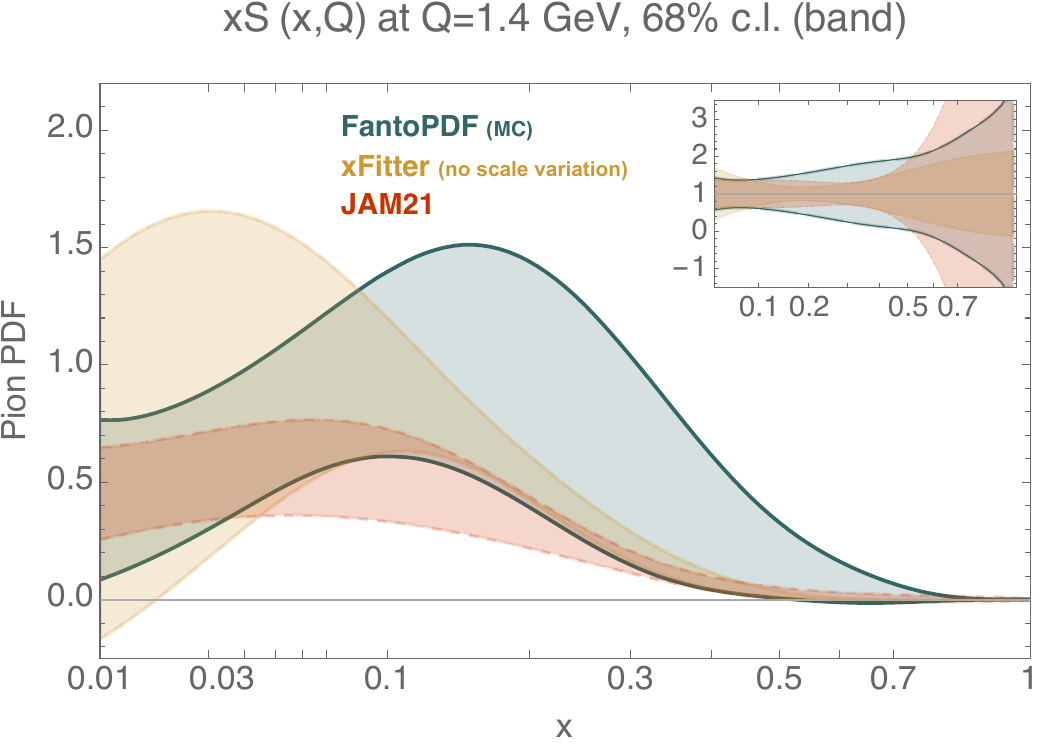}
\includegraphics[width = .95\columnwidth]{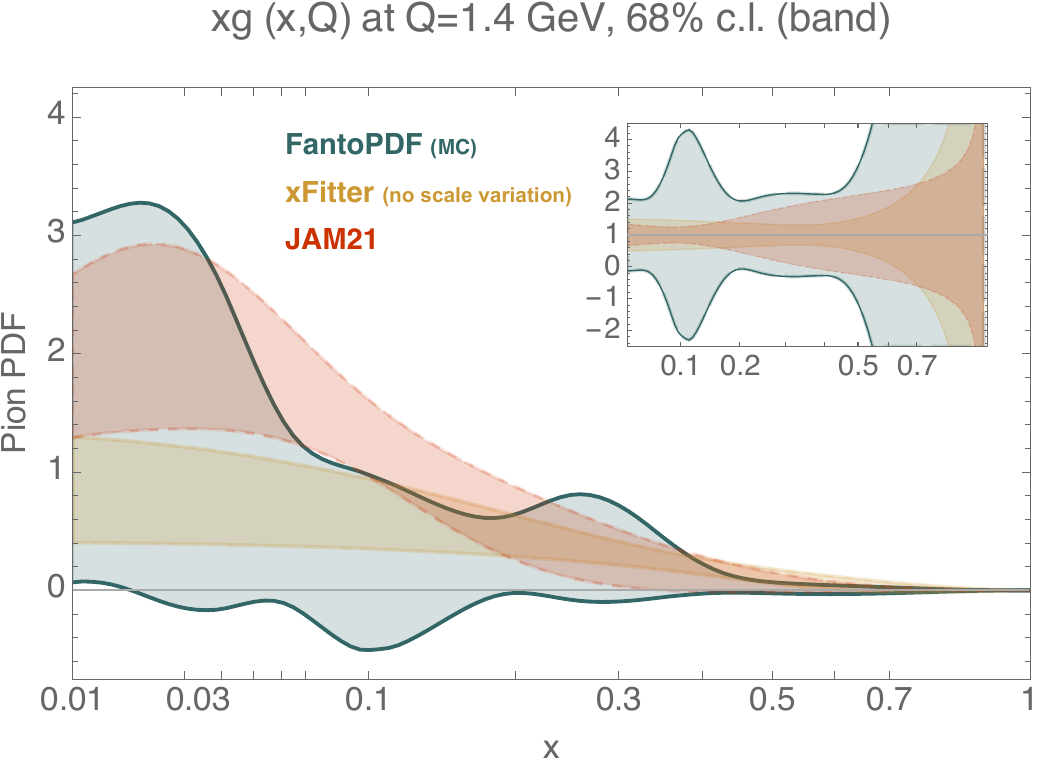}
\caption{Same as in Fig.~\ref{fig:fin_v}, for the quark sea PDF (top), gluon PDF (bottom).}
\label{fig:fin_sg}
\end{center}
\end{figure}
JAM'18~\cite{Barry:2018ort}  reports a best fit of $\chi^2/N_{\rm
  pts}{=}0.98\ (244.8/250)$, with this number being obtained with a specific
  prescription for the pion flux in Eq.~(\ref{eq:pionSF}). 
  JAM'21~\cite{Barry:2021osv} adds large-$x$ resummation corrections in cross sections 
  for the DY data; however, their baseline fit achieves 
$\chi^2/N_{\rm pts}=0.81$.   The double Mellin method for the resummation marginally
improves this result, while other resummation methods worsen it. 
JAM's low $\chi^2$ value might result from the specific selection of
cuts and models for the data. [The objective of the \fanto analysis was
  not to explore the model dependence within the LN data].  
On the other hand, the \xfitter pion fit has achieved 
$\chi^2/N_{\rm pts}=1.17\ (444/379)$, or 1.19 with respect to the number of degrees
of freedom quoted in~\cite{Novikov:2020snp}.

Turning our attention to the shape of the distributions, we observe
that, for the valence sector, the \fanto PDF is in a good overall
agreement with both JAM'21 and \xfitter's results, while displaying a
larger uncertainty in the mid- and especially small-$x$ regions
(Fig.~\ref{fig:fin_v}). The existing DY data do not provide sufficient separation 
between the quark and antiquark PDFs at $x<0.2$, which spectacularly increases the uncertainty on $V(x,Q_0)$ at small $x$ 
compared to the fixed-parametrization fits (in fact, allowing $V(x,Q_0)$ to go negative in some baseline fits, cf. the upper panel of Fig.~\ref{fig:funky_sg}). The $x\to 1$ limit of the \fanto ensemble is
slightly narrower than for JAM'21, possibly because the \fanto's
result was obtained by fitting the unfluctuated data, as opposed
to the bootstrapped one by JAM. 

 The JAM and \xfitter error bands in Fig.~\ref{fig:fin_v} demonstrate another common artifact of fixed polynomial forms -- a pinch point at $x\approx 0.15$, where the uncertainty of $xV(x,Q)$ spuriously vanishes. A pinch point may emerge at some point due to a too restrictive parametric model. The pinch point disappears in the \fanto error band, which combines multiple parametric forms.

\begin{table*}[bth]
\def\hs{\hspace{3.5mm}} %
\begin{tabular}%
    {{|l | c | C{4.cm} C{2.5cm} C{2.5cm}|}} %
    \hline
    \hline
    \rule{0pt}{3ex}
 Name  & $Q$  [GeV]  & \(\langle xV \rangle\)& \(\langle xS \rangle\)& \(\langle xg \rangle\) \\ [0.5ex] 
 \hline\hline      
\rule{0pt}{4ex} 
FantoPDF (DY+$\gamma$+LN) & $\sqrt{1.9}$ & $0.49(8)$ & $0.34(19)$ & $0.18(12)$ \\
\rule{0pt}{4ex} 
xFitter~\cite{Novikov:2020snp} (DY+$\gamma$) & $\sqrt{1.9}$ & $0.55(6)$ & $0.26(15)$ & $0.19(16)$ \\
\rule{0pt}{4ex} 
xFitter w/o scale variation & $\sqrt{1.9}$ & $0.55(2)$ & $0.26(9)$ & $0.19(9)$ \\
\rule{0pt}{4ex} 
JAM'18~\cite{Barry:2018ort} (DY) &$1.27$ & $0.60(1)$ & $0.30(5)$ & $0.10(5)$ \\
\rule{0pt}{4ex}
JAM'18~\cite{Barry:2018ort} (DY+LN)& $1.27$ & $0.54(1)$ & $0.16(2)$ & $0.30(2)$ \\
\rule{0pt}{4ex}
JAM'21~\cite{Barry:2021osv} (DY+LN)& $1.27$ & $0.53(2)$ & $0.14(4)$ & $0.34(6)$ \\
\rule{0pt}{4ex}
JAM'21~\cite{Barry:2021osv} (DY+LN) &&& &\\
\rule{0pt}{1ex}
\hspace{1.cm}
+NLL double Mellin & $1.27$ & $0.46(3)$ & $0.15(7)$ & $0.40(5)$ \\
\hline
\rule{0pt}{4ex}
CT18 NLO (proton)& $\sqrt{1.9}$ & $0.443 (6)$ & $0.160 (10)$ & $0.396(10)$ \\
\rule{0pt}{4ex}
CT18 NNLO (proton)& $\sqrt{1.9}$ & $0.451 (5)$ & $0.157 (10)$ & $0.390 (10)$ \\
  \hline    
\end{tabular}
\caption{
 Pion (rounded) momentum fractions from  global analyses (including this work) at
 $68\%$ C.L., at $Q_0=\sqrt{1.9}$ GeV for the \xfitter-based
 frameworks and $Q_0=1.27$ GeV for JAM. The \FantoPDF results quote
 the MC central values and symmetric errors. The 7th and 8th rows quote JAM'21
 results with, respectively, the baseline and  their preferred large-$x$ resummation technique. The last two rows quote the momentum fractions for the equivalent PDF flavors in the proton from CT18 fits \cite{Hou:2019efy}.}   
   \label{tab:mmts_q0}
\end{table*}

The sea and gluon PDFs go hand-in-hand, as they are tied by DGLAP
evolution as well as by the momentum sum rule constraint. The sea
\FantoPDF (Fig.~\ref{fig:fin_sg}, upper panel) exhibits an extremum
around $x=0.15$ for $Q=Q_0$  that is not fully covered by either JAM
or \xfitter uncertainties. The \fanto uncertainty increases towards
smaller values of $x$, encompassing JAM'21. [The \xfitter's uncertainty for
the sea  is larger at $x> 0.5$ as a consequence of their different 
Hessian diagonalization routine, as discussed in Section~\ref{sec:BezierPion}.]
The \fanto gluon is enlarged across the whole span of $x$. 
The inner frames of Figs.~\ref{fig:fin_v} $\&$~\ref{fig:fin_sg} show the ratios to the central PDF of the corresponding sets.
The mid-$x$ range might be sensitive to the choice of the model for the
pion flux in the description of the LN data. Also, it corresponds to
the transition from the LN and the
pion-induced DY data set, while still being bridged by the prompt-photon
data. The \fanto fit is the first to include all three processes, on
top of the \fanto framework that accounts for uncertainty sources
beyond the aleatory one. 

The first two global analyses of the pion data were performed by the SMRS~\cite{Sutton:1991ay} and GRV~\cite{Gluck:1999xe} groups in the early 90s. At that time, the trend given by the NA3 data on pion-induced Drell-Yan~\cite{NA3:1983ejh} was that the gluon contribution to the momentum fraction was about $47\%$ at $Q^2=4$ GeV$^2$. This fraction was used as a baseline to improve the determination of the sea PDF, resulting in a very small sea contribution. SMRS used a simple $(1-x)^{\beta}$ ansatz for both the sea and the gluon and explored the space of allowed momentum fractions for the sea starting from $5$ to $20\%$, while keeping fixed $(\beta_S, \beta_g)$ parameters. They found that a sea momentum fraction as low as $5\%$ badly fits the NA10 data, and that the $\chi^2$ improves when the sea fraction is increased. On the other hand, GRV adopted a no-sea and valence-like gluon at the initial scale. This scenario similarly leads to a small momentum fraction for the sea, but no further study on the dependence on the model or scans on momentum fraction was done. Both studies were performed at NLO. The values of $\alpha_s(Q^2)$ at that time were given in term of $\Lambda_{\mbox{\tiny QCD}}$ -- a comparison to the modern use of $\alpha_s(M_Z^2)$ might lead to increased differences.

Following upon these early works, in Table~\ref{tab:mmts_q0} we collect the pion momentum fractions from the recent phenomenological fits. %
As was already alluded in Sec.~\ref{sec:results_fanto} in connection with Fig.~\ref{fig:moment_frac_corr}, both \xfitter and \FantoPDF fits prefer a larger $\braket{xS}$ and smaller $\braket{xg}$ as compared to JAM, which primarily reflects the preference of the NA10 data that are different in JAM21 and \xfitter/\fanto fits. (JAM21 uses a set of 56 NA10 data, compared to the \fanto's numbers of data points for NA10 shown in Table~\ref{tab:chi2_break}.) 
 The scenario in which the gluon carries above 40\% of the momentum is disfavored in the \fanto analysis on these grounds, as it results in a too low (anti)quark sea that undershoots the experimental DY cross sections. On the other hand, solutions with a low or even zero gluon improve the description of the DY data and produce low $\chi^2$ values. This conclusion may depend on the order of perturbative QCD calculation and the choice of nuclear PDFs, the factors that will need to be further explored. 

The \fanto uncertainties quoted in Table~\ref{tab:mmts_q0} are significantly larger than from the other groups' fits, reflecting our addition of the parametrization dependence. 
Upon inclusion of the LN DIS data in the \fanto analysis, the ordering of the $\braket{xS}$ and $\braket{xg}$ magnitudes remains the same within the uncertainties. This can be contrasted with the JAM'18 \cite{Barry:2018ort} results in rows 4 and 5 of Table \ref{tab:mmts_q0}, where the ordering of $\braket{xS}=0.30\pm 0.05$ and $\braket{xg}=0.10\pm0.05$ changed to $\braket{xS}=0.16\pm 0.02$ and $\braket{xg}=0.30\pm0.02$ after adding the LN data to the fit.
For the gluon momentum fraction, our estimate for the experimentally allowed interval is $0\leq \braket{xg} \leq 35\%$ at the $2\sigma$ level at $Q_0=\sqrt{1.9}$ GeV. For the quark sea, it is $12\leq \braket{xS} \leq 69\%$, in consistency with the MC histograms in Fig.~\ref{fig:moment_frac}.

\section{Comparison with lattice evaluations}
\label{sec:lattice}

Lattice QCD evaluations are playing an increasingly important role in determining distribution functions from first principles. 
The field is flourishing with the advent of large-momentum effective
theory (LaMET) for lattice evaluations \cite{Ji:2013dva, Ji:2014gla},
followed by alike formalisms such as Ioffe-time distributions (leading
to pseudo-PDFs) \cite{Orginos:2017kos,Radyushkin:2018cvn,Balitsky:2019krf}.    

Several lattice methodologies are available for determination 
of the components of pion PDFs \cite{Lin:2017snn,Constantinou:2020hdm,ExtendedTwistedMass:2021rdx}. 
The first one consists in a computation of the Mellin moments of PDFs. Advanced
 methods include fits of discrete lattice predictions for bare matrix elements of the pion, as well
 as the quasi-PDF approach of LaMET. The quasi-PDF can be related to the
 $P_z$-independent light-cone PDF through a factorization theorem that
 identifies a perturbative matching coefficient with corrections
 suppressed by the hadron momentum. Various groups have evaluated
 pseudo- or quasi-PDFs to NLO~\cite{Zhang:2018nsy, Izubuchi:2019lyk,
   Joo:2019bzr, Gao:2020ito, Sufian:2019bol, Sufian:2020vzb,
   Karpie:2021pap} and well as to NNLO~\cite{Gao:2022iex,
   Zhang:2023bxs} using perturbative matching between the Euclidean
 time quantities and the light-cone PDF. Predictions for the gluon PDF in the
 pion and its momentum fraction are available in Refs.~\cite{Shanahan:2018pib,Fan:2021bcr,Good:2023ecp}. 
 
\begin{table}[bth]
\begin{tabular}%
    {{|l | c | c |c |}} %
    \hline

    \hline
    \rule{0pt}{3ex}

  Name  &  $Q$[GeV]  & \(\langle x (u+\bar{u})_{\pi^+} \rangle\)& \(\langle x g \rangle\) \\ [0.5ex] 
 \hline\hline      

\rule{0pt}{4ex} 

FantoPDF  & $2$ & $0.331 (25)$ & $0.24(10)$ \\

  \hline
    \hline
\rule{0pt}{4ex}
HadStruct~\cite{Joo:2019bzr} & $2$ & $0.2541(33)$ & -- \\
\rule{0pt}{4ex}
Ref.~\cite{Gao:2020ito}  & $3.2$ & $0.216(19) (8)$  & -- \\
\rule{0pt}{4ex}
ETM~\cite{Alexandrou:2020gxs}& $2$ & $0.261(3) (6)$ & -- \\
\rule{0pt}{4ex}
ETM~\cite{ExtendedTwistedMass:2021rdx}& $2$ & $\left.0.601 (28)\right|_{u+d}$ & $0.52 (11)$ \\
\rule{0pt}{4ex}
Ref.~\cite{Meyer:2007tm}& $2$ & -- & $ 0.37(8)(12)$  \\ 
\rule{0pt}{4ex}
Ref.~\cite{Shanahan:2018pib}& $2$ & -- & $ 0.61(9)$  \\ 
\rule{0pt}{4ex}
Ref.~\cite{Good:2023ecp}& $2$ & -- & $ 0.364(38)(36)$  \\
  \hline
\rule{0pt}{4ex}
ZeRo Coll.~\cite{Guagnelli:2004ga}& $2$ & $0.245(15) $ & --\\
\rule{0pt}{4ex}
Ref.~\cite{Martinelli:1987zd}& $7$ &  $0.02$ & -- \\
 \hline              
\end{tabular}
\caption{
 Momentum fractions from the \FantoPDF global analysis,
 PDF-oriented lattice analyses, and early lattice evaluations. Lattice
 collaborations give the combination $u_+=v/2+ 2\bar{u}$ for the $\pi^+$ for even (``odd" in lattice nomenclature) moments. Lattice uncertainties are quoted as ``statistical" and ``systematic," respectively.}   
   \label{tab:mmts}
\end{table}

\begin{figure}[bht]
\begin{center}
    \includegraphics[width=0.995\columnwidth]{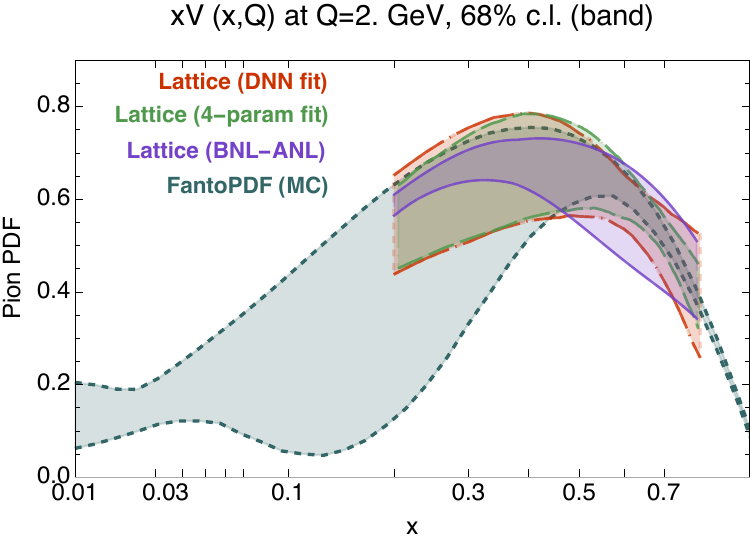}
    \caption{ Valence pion PDF  $x V(x, 2 $ GeV$)$ for the \FantoPDF in dark cyan, as well as for the lattice result of~\cite{Gao:2022iex} in red (DNN version) and green (4-parameter version), and BNL-ANL21~\cite{Gao:2021dbh} in purple. }
    \label{fig:lattice_val}
\end{center}
\end{figure}

In the pion sector, the lattice data become predictive enough to complement the experimental data that are still mainly based on the pion-induced DY process. Lattice predictions may even circumvent the complex theoretical aspects that arise in phenomenological fits when the pion PDF is determined in perturbative QCD at a relatively low scale $Q_0$~\cite{Courtoy:2020fex}. The current lattice predictions for PDFs are most trusted at $0.2\lesssim x \lesssim 0.8$, although their extension to even larger $x$ can be informative given some contention about the behavior of the pion PDFs in the $x\to 1$ limit.
For the valence PDF, lattice groups have found the effective large-$x$ exponent \cite{Ball:2016spl,Courtoy:2020fex}
in the range $0.5\lesssim\beta[=C_V^{\mbox{\tiny eff}}]\lesssim 2$ at a low
scale, showing a strong dependence on the choice of methodology by each lattice group,
including handling of the inverse problem and fits, as discussed in~\cite{Gao:2020ito, Cichy:2021lih}. The NNLO results of
Ref.~\cite{Gao:2022iex} show compatibility of several lattice representations
of the pion PDF with $\beta[=C_V^{\mbox{\tiny eff}}]=1$. As discussed in
Sec.~\ref{sec:results_fanto}, our analysis is in agreement with
the previous fixed-order extractions~\cite{Barry:2018ort, Novikov:2020snp}
as well as those lattice determinations that obtain $\beta=1$. 
Notice,
however, that the uncertainty at mid-$x$ values is larger on those
lattice results, as a result of the corresponding fitting procedure
and the lack of accuracy towards smaller $x$ values. In particular,
the \fanto valence PDF, with its wider uncertainty for the valence
region (Fig.~\ref{fig:combi_CV}), is compatible with the NNLO lattice
pion PDF~\cite{Gao:2022iex}, as illustrated in Fig.~\ref{fig:lattice_val}. 
Recently, Ref.~\cite{JeffersonLabAngularMomentumJAM:2022aix} put forth 
a hybrid determination using both raw lattice data for
reduced Ioffe time pseudo-distributions and current-current
correlator, as well as experimental data. 
Results were found for the valence PDF, and no significant increase in
$C_V^{\mbox{\tiny eff}}$ is reported. 
\begin{figure}[tb]
\begin{center}
    \includegraphics[width=0.995\columnwidth]{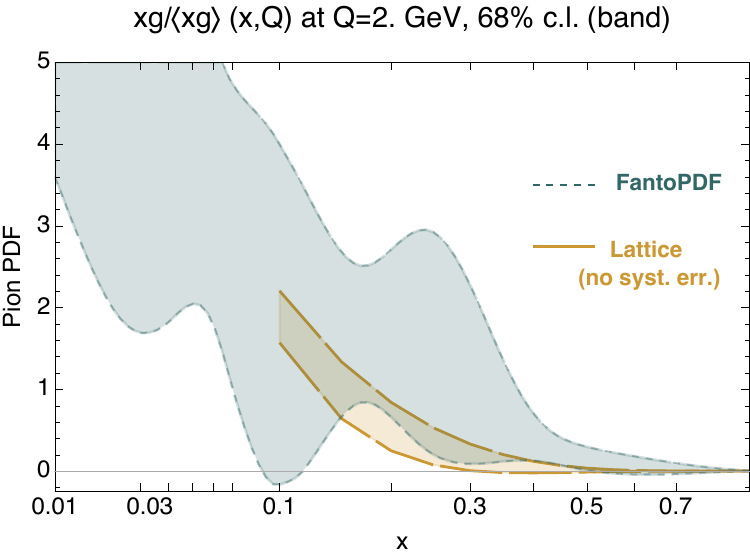}
    \caption{ Gluon pion PDF normalized to its momentum fraction $x g(x, 2 $ GeV$)/\langle x g(x, 2 $ GeV$)\rangle$ for the \FantoPDF in dark cyan and for the lattice result of~\cite{Fan:2021bcr} in yellow mustard. Only statistical uncertainties are quoted for the lattice prediction. }
    \label{fig:lattice_gluon}
\end{center}
\end{figure}

The lattice estimate of the gluon PDF~\cite{Fan:2021bcr} shows a large
dependence on the choice of functional form for $x<0.5$, where the
lattice extrapolation region starts. \fanto results are compatible
with that lattice evaluation at large $x$ [see
  Fig.~\ref{fig:lattice_gluon}], though we cannot conclude on the true
uncertainty of the lattice determination. The lattice-predicted PDF
is normalized to its momentum fraction, leaving the determination of
the overall moment aside, and does not include potentially significant systematic uncertainties. That analysis has been recently complemented  by evaluating the gluon momentum fraction~\cite{Good:2023ecp}. The authors state that the resulting, absolute, gluon PDF is in agreement with previous global analyses. 
\\

In Table~\ref{tab:mmts}, we also compare our results for the second
moments of the \fanto pion PDF with some lattice results for the $q_+$ combination. On the
lattice, one can compute odd moments (corresponding to even powers of $x$ in the integrand) for the ``minus" non-singlet distributions,  $q-\bar{q}$, and even moments (corresponding to odd powers) for the ``plus" distributions.
Due to its large sea component, the \fanto ensemble's second-Mellin moment overshoots the 
current lattice estimations. 
 In the fourth column, we compare the momentum fraction of the \fanto  PDF for
the gluon with some lattice estimates. %
Only one lattice evaluation decomposes the pion momentum into its gluon and quark contributions~\cite{Alexandrou:2021mmi}. For other lattice evaluations of the gluon contributions, the quoted momentum fractions, obtained for  values
$m_{\pi}^{\rm latt} \gg m_{\pi}^{\rm phys}$, are rather large.
The momentum fraction
$\braket{xg}=0.61\pm0.09$ obtained from the analysis of the gravitational form
factor~\cite{Shanahan:2018pib} leaves very little room for the momenta of the valence
and sea distributions. In that sense, it is hardly compatible with the state-of-the-art 
valence momentum fractions on the lattice, nor with the \fanto output, despite the large uncertainty
of the latter.

Regarding the reconstruction of the $x$-dependence of the pion valence PDF from lattice moments, we evaluate the odd moments for $V^{\pi^+}/2$ and compare to the results obtained in Ref.~\cite{Gao:2022iex}, see Table~\ref{tab:mmts_odd}. The third and fifth moments are in good agreement, while the seventh is higher for the  \FantoPDF set [similar results are obtained with JAM21, see Fig.~10 of \cite{Gao:2022iex}]. The behavior of the valence quarks from the \fanto methodology agrees with lattice determinations up to high moments.

\begin{table}
    \begin{tabular}{ | l | l | l | l |}
    \hline
    \hline
    \rule{0pt}{3ex}
    Name & $\langle x^2 V/2\rangle$ & $\langle x^4 V/2\rangle$ & $\langle x^6 V/2\rangle$
    \\ \hline\hline 
    \rule{0pt}{3ex}
    \FantoPDF & $0.110(10)$ & $0.040(2)$ & $0.021(1)$ \\
    \hline
    \rule{0pt}{3ex}
    Ref.~\cite{Gao:2022iex} & $0.1104(73)(48)$ & $0.0388(46)(57)$ & $0.0118(48)(48)$  \\ 
    \hline
    \end{tabular}
    \caption{ Odd Mellin moments compared to the lattice evaluation of Ref.~\cite{Gao:2022iex}, at 2 GeV. 
    }   
   \label{tab:mmts_odd}
\end{table}

\section{Conclusions \label{sec:Conclusions}}
In this study, we extend the informative phenomenological NLO analysis \cite{Novikov:2020snp} 
of charged pion PDFs implemented in the \xfitter program to include the PDF parametrization dependence -- an important factor in the total uncertainty on PDFs that was not considered by previous pion studies. The magnitude of uncertainty tolerated by experimental measurements affects comparisons with predictions for PDFs by nonperturbative QCD approaches. Reliable quantification of uncertainties becomes pivotal in light of the encouraging new developments both in theoretical simulations and experimental measurements of the meson structure. 

The essence of the phenomenological global analyses consists in finding a range of 
functional forms for the PDFs that are compatible with the experimental data. Any such parton distribution can be approximated by polynomials in the momentum fraction $x$ at a low scale $Q_0$ as a consequence of the Stone-Weierstrass theorem. Hence, evidence for PDFs translates into a certain functional, and even often polynomial, behavior. As argued in Ref.~\cite{Courtoy:2020fex}, the property of {\it
  polynomial mimicry} hinders a unique ``if and only if" discrimination among the functional behaviors. 
Mimicry implies the equivalence in describing the same discrete data 
by multiple polynomial shapes of (often) differing orders in interpolations.  
In global analyses for which an initial assumption about the allowed functional forms is necessary, 
the choice of a polynomial shape contributes to the epistemic uncertainty of the final PDF ensemble. This uncertainty plays an increasing role with growing precision of measurements ~\cite{Kovarik:2019xvh,Hou:2019efy,Courtoy:2022ocu} and must reflect generic prior considerations such as positivity \cite{Candido:2020yat, Collins:2021vke, Candido:2023ujx} or flavor structure in asymptotic limits \cite{Hou:2019efy,Accardi:2023gyr}.  

Interestingly, the uncertainty quantification in the context of
polynomial shapes is also of concern to any analysis facing an inverse
problem; it has been discussed for lattice determinations of, e.g.,
the shape of the valence pion PDF~\cite{Gao:2020ito, Cichy:2021lih}. 
Nonparametric methodologies for QCD global analyses, such as those based
on ML/AI, are not free from that source of uncertainty. One way to address the epistemic uncertainty
is to estimate it by representative sampling of acceptable functional forms~\cite{Courtoy:2022ocu}.  
The open-ended inverse problem of finding the epistemic uncertainty is also ill-posed, yet sufficiently dense sampling over the functional forms provides at least a lower estimate with well-designed, parsimonious parametric models.

\begin{table*}[t]
  \centering
  \begin{tabular}{|C{1.75cm}|C{1.75cm}|C{2.25cm}C{2.25cm}C{2.25cm}C{2.25cm}C{2.25cm}|C{1.75cm}|}
\hline 
$\langle xg(Q_{0})\rangle$ & $\langle xS(Q_{0})\rangle$ & E615 & NA10-194 & NA10-286 & WA70 & HERA $F_{2}^{\pi}$ & $\chi_{tot}^{2}$\tabularnewline
 &  & (140) & (67) & (73) & (99) & (29) & (408) \tabularnewline
\hline 
\hline 
\rule{0pt}{3ex} 0.32 & 0.14 & 204.7 & 107.7 & 101.5 & 25.6 & 4.7 & 450.1\tabularnewline
\hline 
\rule{0pt}{3ex} 0.27 & 0.17 & 205.5 & 106.4 & 101.5 & 28.8 & 4.3 & 450.8\tabularnewline
\hline 
\rule{0pt}{3ex} 0.20 & 0.28 & 207.5 & 100.7 & 102.5 & 27.0 & 7.1 & 447.6\tabularnewline
\hline 
\rule{0pt}{3ex} 0.12 & 0.45 & 208.4 & 99.6 & 94.9 & 31.6 & 2.1 & 440.4\tabularnewline
\hline 
\rule{0pt}{3ex} 0 & 0.64 & 204.6 & 95.8 & 93.6 & 29.6 & 4.94 & 432.2\tabularnewline
\hline 
\end{tabular}

  \caption{Gluon and sea momentum fractions, and $\chi^2$ contributions from each experimental set for the five fits combined in \FantoPDF. The number of data points for each set is given in the parentheses. The fits are listed in the order of the decreasing gluon momentum fraction $\langle xg (Q_0) \rangle$.}
  \label{tab:chi2_break}
\end{table*}

In this manuscript, we explore the positive aspect of the polynomial
mimicry. In Ref.~\cite{Courtoy:2020fex}, we contemplated the possibility of using Bézier
curves -- linear combinations of Bernstein polynomials of a user-specified polynomial degree -- as continuous 
approximations of PDFs determined by the PDF values at a few chosen points in $x$. The algorithmic nature of such construction makes it versatile yet more transparent than the complementary approaches based on neural networks. Indeed, a Bézier parametrization is directly defined by the PDF values at control points 
and not by latent parameters and hyperparameters of a neural network. 
We introduced \fanto, a C++ program providing 
\meta parametrizations for PDFs that capture the asymptotic limits at $x\to 0$ and $x \to 1$
using the carrier component and any features of PDFs at intermediate $x$ using the modulator component. 
The modulator function includes a Bézier curve of degree $N_m$ with coefficients determined by 
$N_m+1$ control points that are optimally placed according to the problem to solve. 
When solving the inverse problem for PDFs, minimization of the objective function varies the free parameters
of the carrier function and values of the modulator at the control points. 
A variety of functional forms can be generated on the fly by using different carriers, placements of fixed control points, and the $x$-stretching variable in the modulator.

This \meta parametrization module has been implemented into the \xfitter QCD fitting framework and tested in the pion analysis at NLO that has been already implemented in ~\cite{Novikov:2020snp}. The latter analysis complements the panorama offered by the three JAM fits (collinear~\cite{Barry:2018ort}, with the inclusion of large-$x$
resummation~\cite{Barry:2021osv}, and simultaneous collinear and TMD
fitting~\cite{Cao:2021aci}) and expands on the pioneering QCD analyses of the pion PDFs in \cite{Sutton:1991ay,Gluck:1991ey,Gluck:1999xe}.
The \xfitter analysis uses the Hessian
methodology \cite{Pumplin:2001ct,Pumplin:2002vw} to estimate aleatory uncertainties associated with the data. To the aleatory uncertainties, we add the epistemic ones obtained by systematic sampling of about 100 modulator forms with degree up to $N_m=3$. We find that the range of pion PDFs rendering essentially the same $\chi^2$ is significantly increased by considering multiple functional forms. 

The most constraining set of the experimental data, coming from the pion-induced Drell-Yan pair production on a tungsten target, is characterized by large momentum fractions for the pion beam. Even after inclusion of the leading-neutron DIS data constraining the gluon PDF down to $x\sim 10^{-3}$, the available experimental measurements only weakly separate the quark and antiquark PDFs, and the quark sea and gluon PDFs at $x < 0.2$. 
The consequence is the enlarged uncertainty bands in the \fanto PDF ensemble seen in Figs.~\ref{fig:fin_v} and \ref{fig:fin_sg}, as well as the extended range of the correlated momentum fractions for the quark sea and gluon in Fig.~\ref{fig:moment_frac_corr}. The \fanto analysis finds that $\braket{xg}$ in the range from 0 to 35\% are allowed by the data.\footnote{For example, a solution with a zero gluon at $Q_0$ in Fig.~\ref{fig:fin_sg} yields a somewhat lower $\chi^2=432.2$ than the other highlighted fits.} This is a wider range than what was found with fixed parametrization forms. 

A curious tension therefore emerges between the small momentum fractions $\braket{xg}$ that may in fact be preferred by the NA10 DY data (which is more extensive in \xfitter than in the JAM analysis), and the first lattice QCD predictions suggesting $\braket{xg}$ above 30\%, as summarized in Table~\ref{tab:mmts}. 
These observations raise the value of the ongoing and planned pion scattering experiments at  AMBER \cite{AndrieuxSPIN} and EIC \cite{Aguilar:2019teb, Arrington:2021biu}. The dependence of DY constraints on the assumed nuclear PDFs needs to be further explored, and additionally, this and other \cite{Novikov:2020snp,
  Barry:2018ort, Barry:2021osv} phenomenological studies are still dominated by the low-$Q$ data for
which higher-order perturbative QCD uncertainties need to be reduced.  We did not add the uncertainty due to the QCD scale variations -- its magnitude can be gauged from the comparison of the \xfitter momentum fractions with and without the scale-variation uncertainty in Table~\ref{tab:mmts_q0}.
Besides the PDFs, 
description of some pion observables by the nonperturbative
theoretical techniques and transition to the perturbative regime is still insufficiently
understood~\cite{Polyakov:2009je,Radyushkin:2009zg,Noguera:2010fe}. The pion structure will thus remain an exemplar to develop QCD methods. Our \xfitter analysis will be made publicly available together with the \fanto module and LHAPDF grids for the NLO pion error PDFs that we produced.

\begin{figure*}[tb]
\begin{center}
    \includegraphics[width=0.49\textwidth]{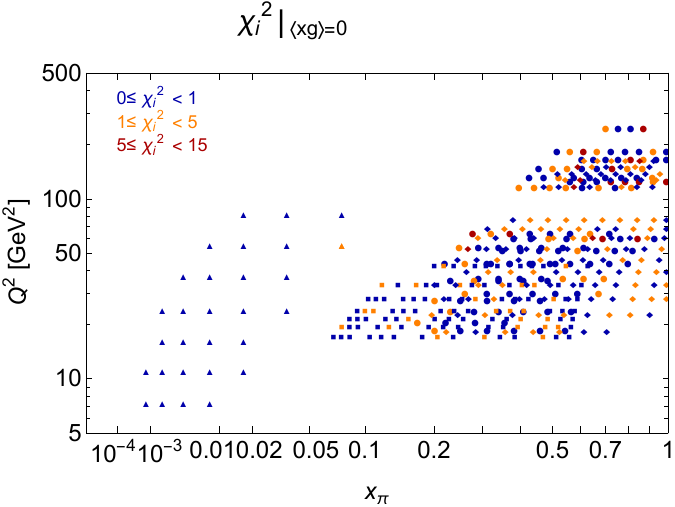}\quad
    \includegraphics[width=0.49\textwidth]{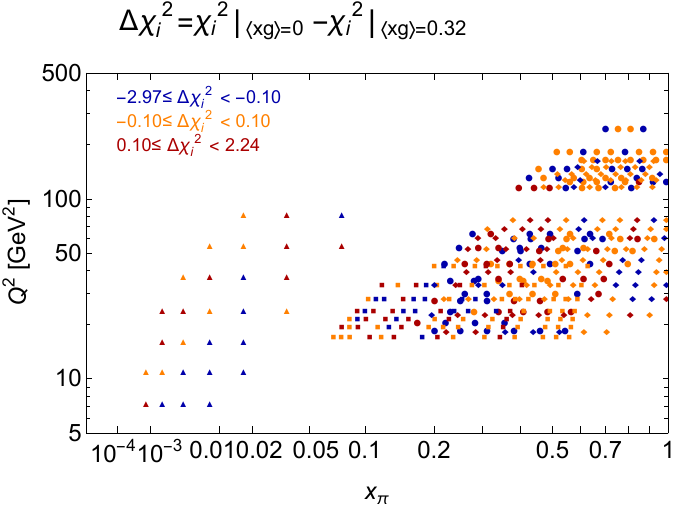}\\
    (a) \hspace{0.49\textwidth} (b)

    \caption{Contributions of individual data points to $\chi^2$ of respective experiments, shown in the plane of $x_\pi$ and $Q^2$ and using the same shapes of symbols for the individual experiments as in Fig.~\ref{fig:xQ2} (circle for NA10, square for WA70, rhombus E615, triangle for HERA leading-neutron data). (a) The color code indicates the partial $\chi^2$ value for each data point for the zero-gluon scenario,  $\langle xg(Q_0^2)\rangle=0$, according to the $\chi^2$ intervals indicated in the legend. (b) The color indicates the differences in the partial $\chi^2$ of the points in the scenarios with $\langle xg(Q_0^2)\rangle=0$ and $\langle xg(Q_0^2)\rangle=0.32$. Most data points show a negligible difference in $\chi^2$ between the two scenarios (orange points).}
    \label{fig:kin_map_lowhigh}
\end{center}
\end{figure*}

\section*{Acknowledgments}

We thank the  \xfitter collaboration members for their assistance with the \xfitter program,
and we additionally thank  I.~Novikov and A.~Glazov for communications on 
the uncertainties of the momentum fractions in the original \xfitter paper.
We are grateful to T.~J.~Hobbs for discussions of the leading-neutron DIS data as well as for arranging 
the collaborative meeting on this project at the Illinois Institute of Technology, and to  K. Mohan for a discussion of the Stone-Weierstrass approximation theorem. 
This study has been financially supported by 
the Inter-American Network of Networks of QCD Challenges,
a National Science Foundation AccelNet project, 
by CONACyT--Ciencia de Frontera 2019 No.~51244 (FORDECYT-PRONACES), PIIF ``Interconexiones y sinergias entre la física de
altas energías, la física nuclear y la cosmología" (UNAM),
by the U.S.\ Department of Energy under Grant No.~DE-SC0010129,
and 
by the  U.S.\ Department of Energy, Office of Science, Office of Nuclear Physics, 
within the framework of the Saturated Glue (SURGE) Topical Theory Collaboration.
AC and MPC were further supported by the UNAM Grant No. DGAPA-PAPIIT IN111222.
PMN was partially supported by the Fermilab URA award, using the resources of the Fermi National Accelerator Laboratory (Fermilab), a U.S. Department of Energy, Office of Science, HEP User Facility. Fermilab is managed by Fermi Research Alliance, LLC (FRA), acting under Contract No. DE-AC02-07CH11359.

\appendix

\section{Description of data sets in pion fits \label{sec:AppendixData}}

Section~\ref{sec:results_fanto} posited that description of the data in our analysis is relatively insensitive to the distribution of the pion net momentum among its quark sea and gluon components, as reflected {\it e.g.} in the plot of the respective momentum fractions in 
Fig.~\ref{fig:moment_frac_corr}. As an illustration, 
Table~\ref{tab:chi2_break} presents the breakdown of the $\chi^2$ values in the five baseline fits over the groups of fitted experiments. In the last column, we see that all fits achieve a good overall $\chi^2$ of order 430-450 for 408 data points. The picture is more nuanced when we look at the individual experiments. In Table~\ref{tab:chi2_break}, all three Drell-Yan experiments (E615, NA10-194, and NA10-286) have  elevated $\chi^2/N_{\rm pts}$ of order 1.4. The WA70 and H1 LN data sets have a very low $\chi^2$, reflecting their large uncertainties. The total $\chi^2$ remains good because of the mutual compensation between the higher-than-expected $\chi^2$'s of the DY experiments and low $\chi^2$'s for the other two experiments. 

The total $\chi^2$ improves by up to 20 units when $\langle xg\rangle$ is reduced from 0.32 (the first row in Table~\ref{tab:chi2_break}) to zero (the last row). The two NA10 data sets both prefer a lower $\langle xg\rangle$, i.e. a large $\langle xS\rangle$, cf. the last two rows. The rest of the experiments show only a weak sensitivity to $\langle xg\rangle$ and $\langle xS\rangle$, with their $\chi^2$ remaining almost constant. 

\begin{figure*}[tbp]
\begin{center}
\includegraphics[width = .48\textwidth]{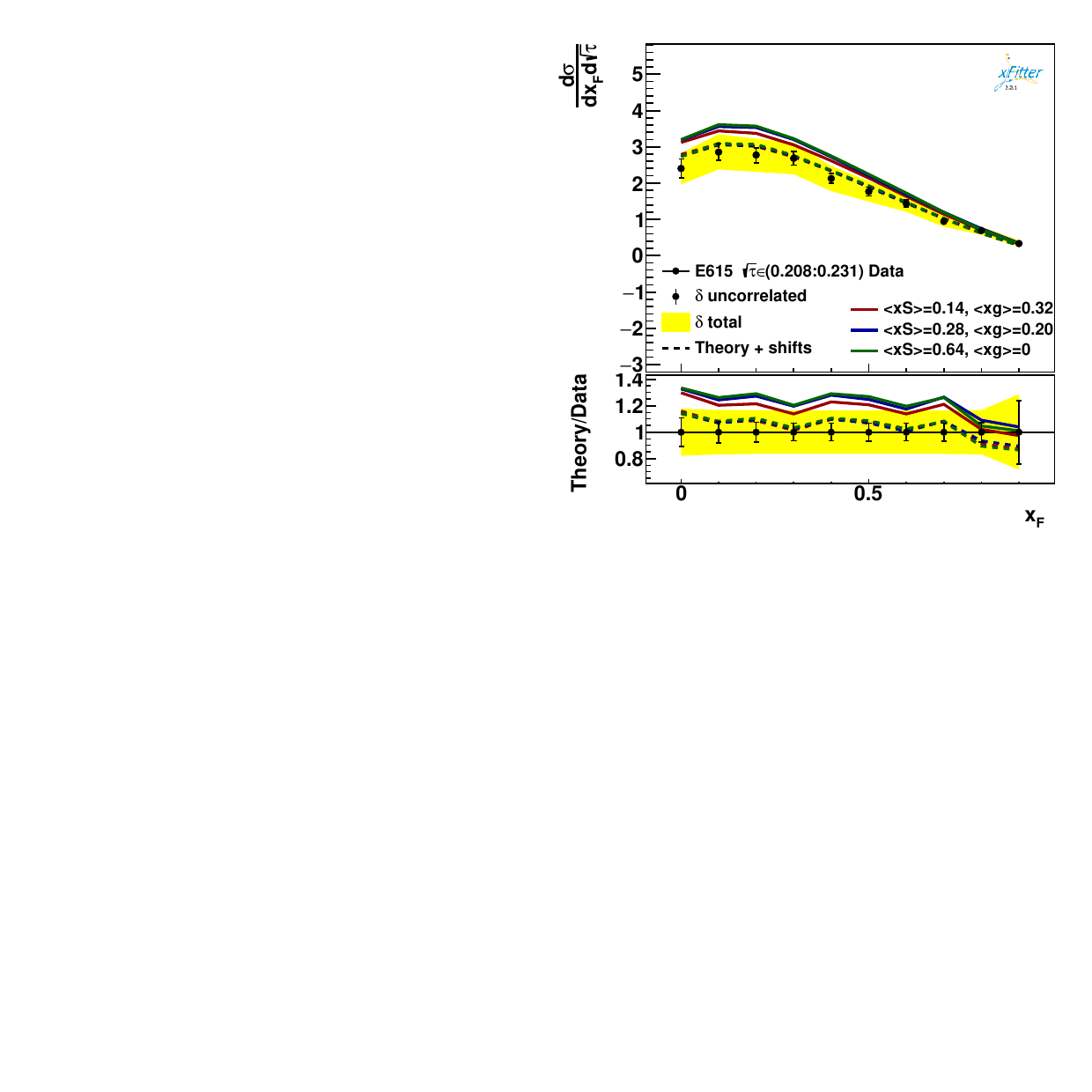}\quad
\includegraphics[width = .48\textwidth]{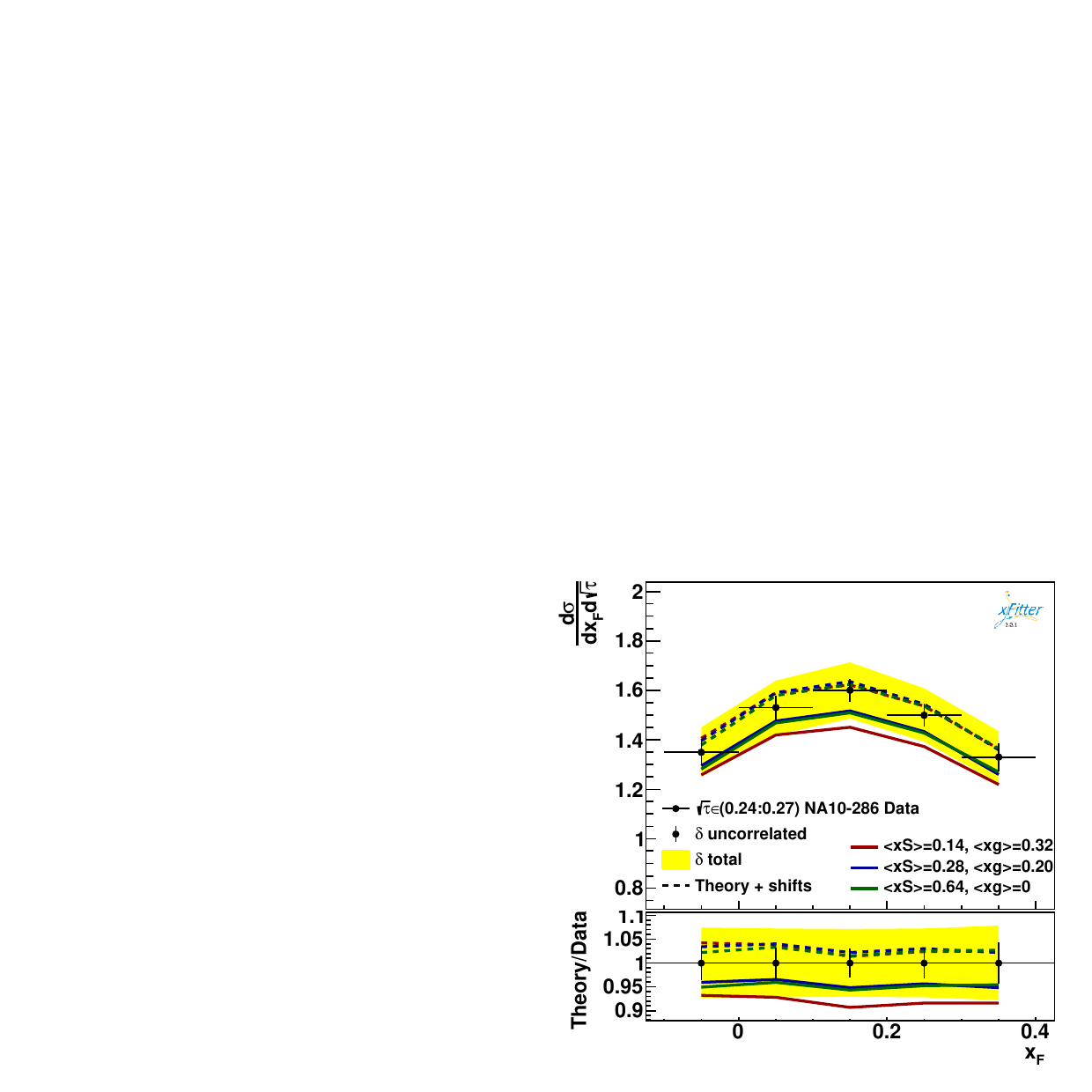}\\
(a)\hspace{.48\textwidth}(b)\\
\includegraphics[width = .48\textwidth]{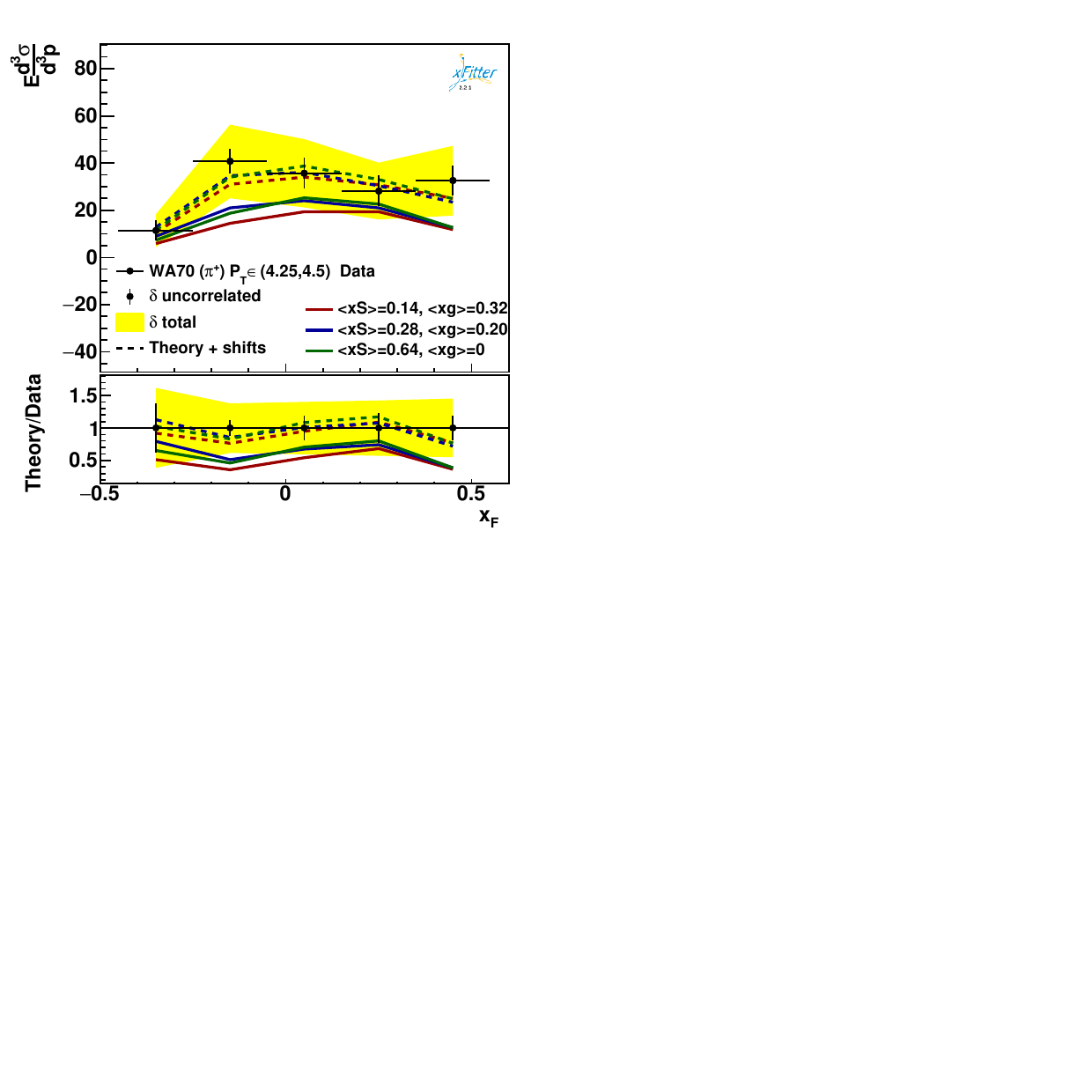}\quad
\includegraphics[width = .48\textwidth]{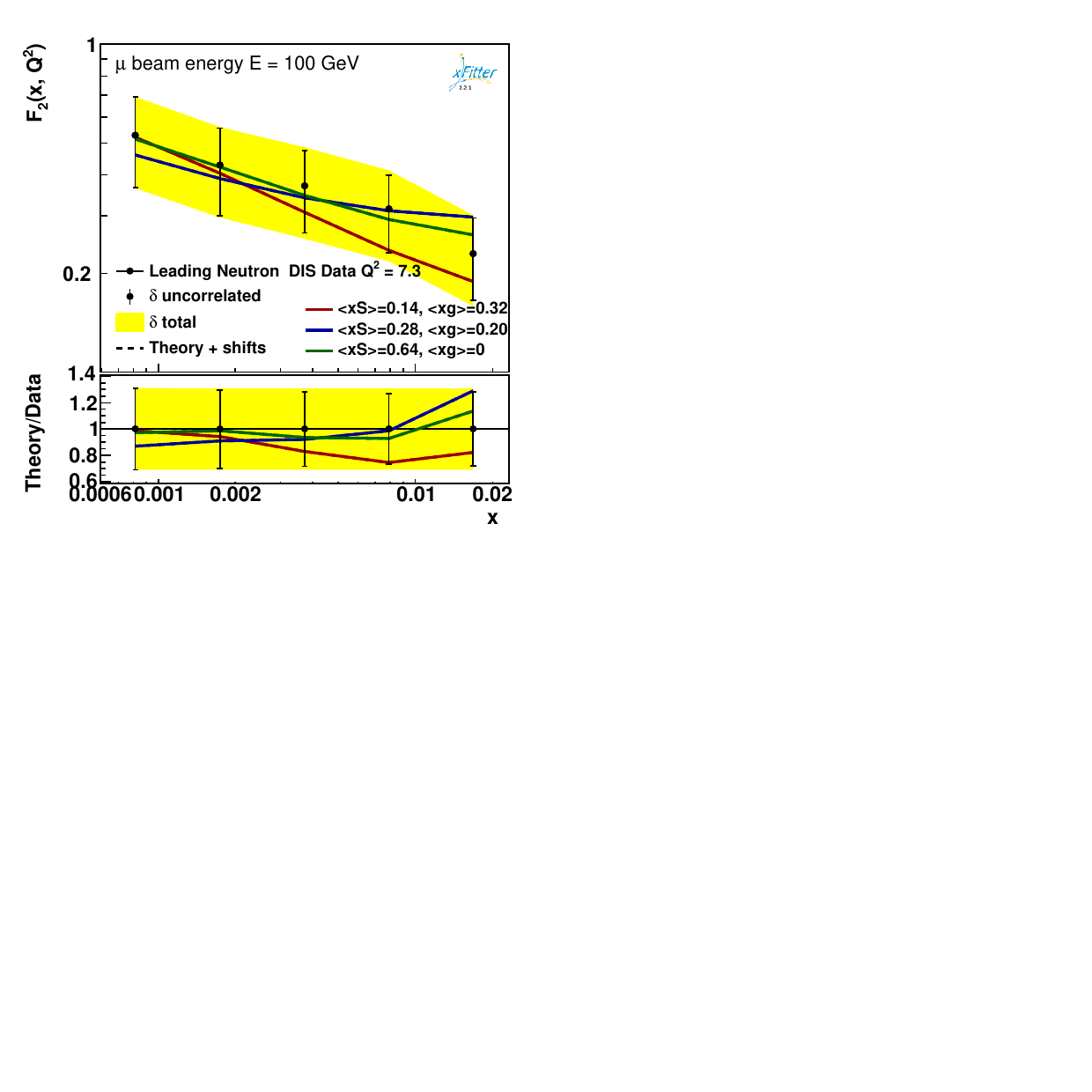}
(c)\hspace{.48\textwidth}(d)
\caption{Comparisons of data and theory (without or with systematic) shifts, 
for typical bins of (a) E615, (b) NA10, (c) WA70, and (d) H1 leading-neutron data sets. The legend indicates the quark sea and gluon momentum fractions for the used PDFs.}
\label{fig:DataExamples1}
\end{center}
\end{figure*}

We can examine contributions of individual data points to elucidate what drives these patterns seen in Table~\ref{tab:chi2_break}. Color-coded kinematic maps in Fig.~\ref{fig:kin_map_lowhigh} quantify how partial $\chi^2$ contributions of the individual data points are distributed in the $\{x_\pi, Q^2\}$ plane that is already familiar from the plot of the fitted data sets in Fig.~\ref{fig:xQ2}. Those maps are particularly illuminating for a large data set, like E615, that spans a substantial kinematic region. 

Fig.~\ref{fig:kin_map_lowhigh}(a) shows the partial $\chi^2$ contributions of the data points in a fit with $\langle xg\rangle=0$ (the same as in Table~\ref{tab:chi2_break}). We see no evidence of localized tensions, since well-fitted and poorly fitted points spread quite uniformly over the kinematic plane. 

Fig.~\ref{fig:kin_map_lowhigh}(b) shows the {\it change} in the partial $\chi^2$'s of the points between the two fits with $\langle xg \rangle =$ 0.32 and 0 from Table~\ref{tab:chi2_break}. 
Once again, there is no clear pattern in these $\chi^2$ variations, confirming degeneracy of the data description with respect to the $\langle xg\rangle-\langle xS\rangle$ correlation. In particular, for the LN data, some points prefer a higher, and other points prefer a lower $\langle xg\rangle$. 

Our final examination focuses on the plots comparing the theory and data bin-by-bin.\footnote{For the leading-neutron data, the $30\%$ uncorrelated uncertainty accounts for the pion flux and normalization uncertainties added in quadrature.} Figure~\ref{fig:DataExamples1} shows some figures of this kind from the collection produced by  
\xfitter for each range of rapidity or $p_T$. For  Fig.~\ref{fig:DataExamples1}, we select a few figures in which one can see the typical patterns present in these comparisons, here shown for  three fits with $\langle xg\rangle =$ 0.32, 0.20, and 0 from Table~\ref{tab:chi2_break}. The solid and dashed lines indicate theory predictions obtained without and with the shifts in the normalization allowed by the experimental systematic uncertainties. Among the Drell-Yan experiments, theory tends to overshoot the data in many bins of the E615 sample and undershoot the data in the NA10 samples. Reducing $\langle xg \rangle$ allows the fit to increase $\langle x S\rangle$ and hence increase the cross section for NA10, leading to a better agreement with the respective data.

In the WA70 direct-photon production, theory also tends to undershoot the data, but nevertheless lies within the large uncertainty bands. The uncertainties of the LN data are still too large to introduce any preference. The overall level of precision in these comparisons is consistent with the modest accuracy (NLO) of the used theoretical predictions, as well as with the underlying uncertainties in the measured data and modeling of nuclear targets.

\bibliographystyle{utphys}
\bibliography{biblio4fantomas}

\end{document}